\documentclass{aa}  

\usepackage{appendix}
\usepackage{graphicx}
\usepackage{txfonts}
\usepackage{dcolumn}
\usepackage{bm}
\usepackage{amsmath}
\usepackage{float}
\usepackage{lipsum}
\usepackage{color}
\usepackage{times}
\usepackage{textcomp}
\usepackage{latexsym}
\usepackage[hidelinks]{hyperref}
\usepackage{acro}
\usepackage{mathtools}
\usepackage{url}
\usepackage[utf8]{inputenc}
\usepackage{physics}
\usepackage{listings}
\usepackage{subfigure}

\newcommand{\vect}[1]{{\bm #1}}

\begin{document}

\title{\textsc{lensingGW}: a \textsc{Python} package for lensing of gravitational waves} 

\author{G. Pagano\inst{1,2}
\and O. A. Hannuksela\inst{3,4}
\and T. G. F. Li\inst{5}
       }
       
\institute{Dipartimento di Fisica ``Enrico Fermi'', Università di Pisa, Pisa I-56127, Italy
\and 
INFN sezione di Pisa, Pisa I-56127, Italy\\
\email{giulia.pagano@pi.infn.it}
\and 
Nikhef -- National Institute for Subatomic Physics, Science Park, 1098 XG Amsterdam, The Netherlands\\
\email{o.hannuksela@nikhef.nl}
\and 
Department of Physics, Utrecht University, Princetonplein 1, 3584 CC Utrecht, The Netherlands
\and 
Department of Physics, The Chinese University of Hong Kong, Shatin, NT, Hong Kong\\
\email{tgfli@cuhk.edu.hk}
        }

\date{\today}

\abstract
{Advanced LIGO and Advanced Virgo could observe the first lensed gravitational waves in the coming years, while the future Einstein Telescope could observe hundreds of lensed events. 
Ground-based gravitational-wave detectors can resolve arrival time differences of the order of the inverse of the observed frequencies. As LIGO/Virgo frequency band spans from a few $\rm Hz$ to a few $ \rm kHz$, the typical  time resolution of current interferometers is of the order of milliseconds. When microlenses are embedded in galaxies or galaxy clusters, lensing can become more prominent and result in observable time delays at LIGO/Virgo frequencies. Therefore, gravitational waves could offer an exciting alternative probe of microlensing. 
However, currently, only a few lensing configurations have been worked out in the context of gravitational-wave lensing. 
In this paper, we present \textsc{lensingGW}, a \textsc{Python} package designed to handle both strong and microlensing of compact binaries and the related gravitational-wave signals. 
This synergy paves the way for systematic parameter space investigations and the detection of arbitrary lens configurations and compact sources. 
We demonstrate the working mechanism of \textsc{lensingGW} and its use to study microlenses embedded in galaxies.}


\maketitle

\section{Introduction} 
\label{sec:introduction}

The Advanced LIGO~\citep{TheLIGOScientific:2014jea,TheLIGOScientific:2016agk} and Advanced Virgo~\citep{TheVirgo:2014hva} gravitational-wave detectors observed ten binary black hole mergers during the observation runs O1 and O2~\citep{PhysRevX.9.031040} and dozens of them during the observation campaign O3~\citep{GCNO3}. 
With the prospects of the additional detectors KAGRA and LIGO India joining the global gravitational wave (GW) network~\citep{somiya:2012detector,aso2013interferometer,akutsu2018construction,M1100296} and with the recently approved A+ detector upgrade~\citep{InstrumentationWhitePaper}, this number is expected to eventually reach hundreds~\citep{2016ApJ...833L...1A,PhysRevX.9.031040}.

As the number of detections grows, several novel avenues will open in the field of GWs~\citep{abbott2018:prospects}. 
One example of such an avenue is \emph{gravitationally lensed} GWs, first of which may be possible to observe within the coming years~\citep{ng2017precise, li2018gravitational, 2018MNRAS.480.3842O}. It has been suggested that lensed GWs could offer interesting applications in fundamental physics, astrophysics, and cosmology~\citep{2011MNRAS.415.2773S, Liao:2017ioi,PhysRevLett.118.091101,PhysRevLett.118.091102, PhysRevD.95.063512, Lai:2018rto,  dai2018detecting, 2019arXiv190808950M, 2019RPPh...82l6901O,10.1093/mnras/staa1430,Jung:2017flg,Cao:2019kgn,Hou:2019dcm,Sun:2019ztn,Hannuksela:2020xor}. 
The first searches for gravitational-wave lensing signatures in the LIGO and Virgo data were carried out recently~\citep{Hannuksela:2020xor,2019arXiv190406020L,2019arXiv191205389M,Singer:2019vjs,10.1093/mnras/staa1430}, finding no clear evidence of lensing, despite discussion of early detections~\citep{broadhurst2018reinterpreting,Broadhurst:2019ijv}.

When gravitational waves propagate near massive astrophysical objects, their trajectories will curve, which can result in gravitational lensing and multiple images. 
As we observe the waves from each of the multiple image, their amplitudes will have changed due to the focusing by lensing, and they will arrive at different times due to having traveled different trajectories at the same speed~\citep{ohanian1974focusing,bliokh1975diffraction,bontz1981diffraction,thorne1983theory,deguchi1986diffraction,nakamura1998gravitational,takahashi2003gravitational,2018MNRAS.480.3842O,broadhurst2018reinterpreting,Broadhurst:2019ijv,Contigiani:2020yyc}. 

However, unlike in the lensing of electromagnetic waves, where one classifies incoming photons as lensed by their angular direction, LIGO/Virgo requires entirely different and novel methodologies to classify GWs as lensed. That is, by statistically distinguishing them as identical events using GW templates and Bayesian analysis~\citep{Haris:2018vmn,hannuksela2019search,2019arXiv190406020L,2019arXiv191205389M}.
As opposed to being limited by the angular resolution of optical detectors, the precision at which we will observe these waves in the interferometers is limited by LIGO/Virgo's millisecond time resolution. 

In contrast to the images produced by galaxies and galaxy clusters (\textit{strong lensing}), the lensing effects induced by masses of roughly $10^{-6}\lesssim M/M_\odot\lesssim 10^6$ are referred to as \textit{microlensing}~\citep{schneiderBook}.
If there are microlenses on the path of strongly lensed waves, the strong lens can enhance the microlensing effect~\citep{diego2018dark,diego2019observational}, resulting in beating patterns that could allow us to infer the lensing object's properties such as its mass (see Refs.~\citep{Lai:2018rto} and~\citep{Christian2018}  for applications to intermediate-mass and stellar-mass black hole/star lenses). 
Indeed, a recent work by Diego et al.~\citep{diego2019observational} demonstrated that strongly lensed GWs from extragalactic sources are more likely to be microlensed, as the effective Einstein radius of the microlenses grows in size thanks to the strongly lensing galaxy.

The inference of such signals relies on the ability to correctly identify the lensed gravitational wave. To model these lensed GWs, one needs to be able to predict the properties of lensed images formed by generic matter distributions.

When microlenses are embedded in a macromodel (a galaxy or a galaxy cluster), \textit{microimages} separated by scales as small as $ \mu \rm{as}$ can form~\citep{diego2019observational}. Indeed, in order to model this interaction, one must consider both strong lensing and microlensing together~\citep{diego2019observational}. Therefore, to model the lensed GW accurately, one has to identify both the strongly lensed images formed by the galaxy/galaxy cluster (whose typical separations are $\lesssim \rm as$~\citep{schneiderBook, collett2015}) and the microimages formed by the microlenses.

Software packages that address the lensing of GWs must then be able to predict images with diverse separations even when no \textit{a priori} knowledge on their range is available. However, existing software programs for gravitational lensing focus on light lensing rather than on gravitational-wave lensing~\citep{lenstronomy, gravlens}. As a consequence, they do not provide lensed GW signals and target strongly lensed images as opposed to microimages. 

In this paper, we present \textsc{lensingGW}, a software package to model lensed images of compact binaries from arbitrary lens models. \textsc{lensingGW} has been designed to handle both the strong lensing and microlensing of GWs simultaneously. Thus, it can determine the strongly lensed images produced by galaxies/galaxy clusters and microimages induced by microlensing backgrounds, with no a priori assumptions on the image structure, while retaining fast performance.
The code is designed to provide GW astronomers and data analysts with a user-friendly tool for the analysis of lensed GW signals. 
Besides supplying the ordinary output of lensing codes, such as image positions, magnifications and time delays, \textsc{lensingGW} also simulates lensed and unlensed GWs and their associated detector strains for generic lensing configurations, including multi-component strong lenses with an arbitrary number of microlenses.

The latter is necessary for assessing the detectability of lensed events through mismatch and signal-to-noise ratio (SNR) calculations~\citep{Allen_2012} and for investigating the parameter space where microlensing becomes relevant. 
Thus, this work will be a step towards establishing a complete lensing framework for GWs, an effort which has seen a recent push from both astronomy and GW modeling sides~\citep{2019MNRAS.485.5180S,2018IAUS..338...98S,2020arXiv200201479R,diego2019observational,cao2014gravitational,Dai:2017huk,dai2017effect,Lai:2018rto,Christian2018,dai2018detecting,Haris:2018vmn,hannuksela2019search,Mehta:2019aa}.

We review the theoretical framework for lensing in Sec.~\ref{sec:lensingtheory} and present our software in  Sec.~\ref{sec:lensinggw}. Results validating our code for simple test cases are illustrated in  Sec.~\ref{sec:validation}. The application of \textsc{leningGW} to scenarios of astrophysical interest is demonstrated in Sec.~\ref{sec:applications}, where we present an example mismatch computation and a more realistic lensing system with microlenses embedded within a galaxy. We conclude in Sec.~\ref{sec:conclusions}.

\section{Lensing of gravitational waves}
\label{sec:lensingtheory}

Lensing modifications to the GW waveform can be solved in the wave optics limit from the Einstein field equations, when the gravitational potential $U$ is too weak to change the polarization of the wave ($U\ll 1$) and when the gravitational wave can be separated from the background space-time~\citep{nakamura1998gravitational,takahashi2003gravitational}\footnote{Note that, when the wavelength of the gravitational wave is much larger than the object's size and the wave travels near the object, the wave may no longer be separated from the background and wave scattering occurs (see, e.g.,~\citep{takahashi2005scattering}).}.

\begin{figure}
\centering
 \includegraphics[width=0.9\columnwidth]{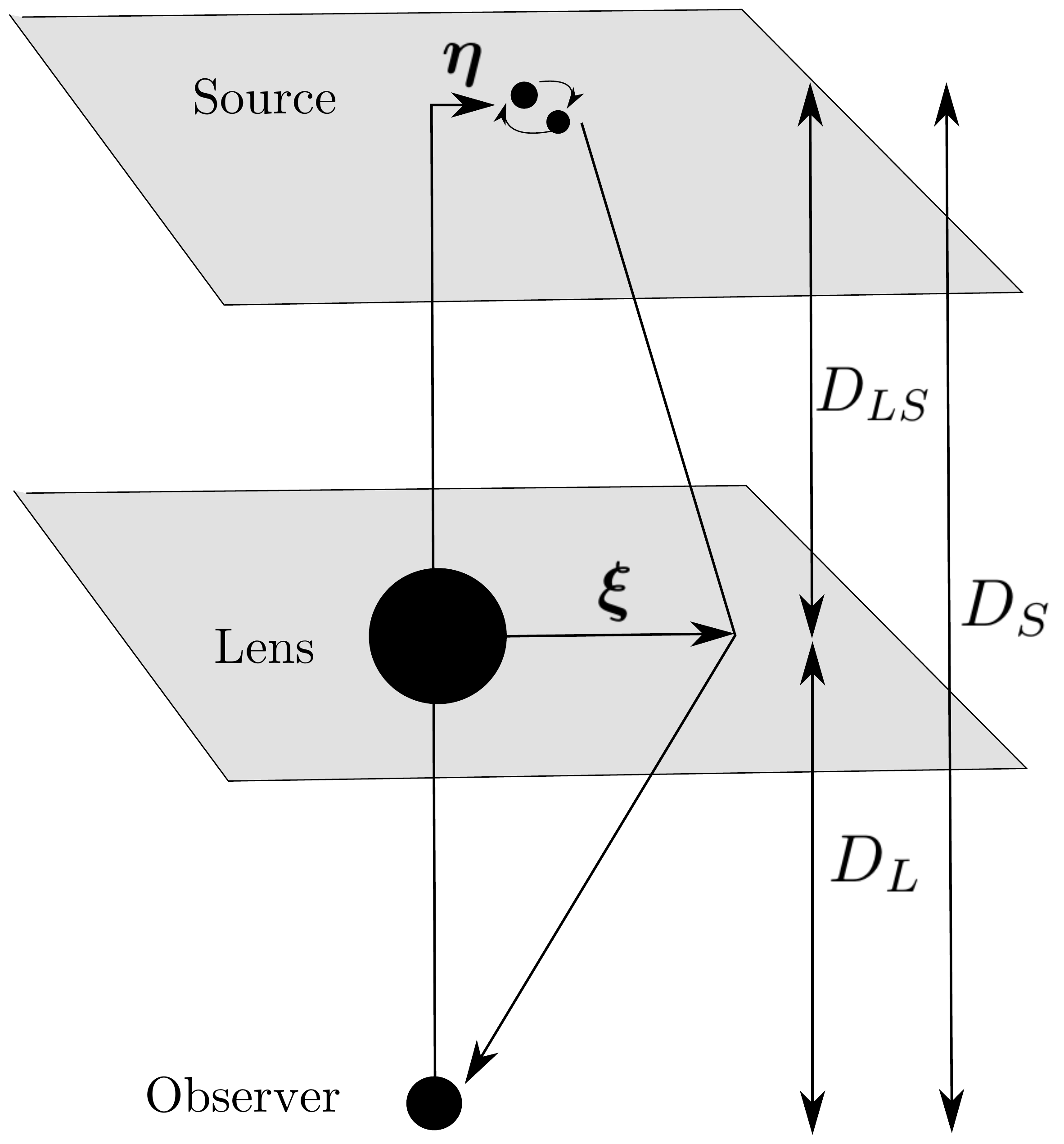}
 \caption{\textit{Illustration of the gravitational lens configuration in the thin lens approximation}. 
 The lensing configuration is described by the source displacement from the line-of-sight $\vect{\eta}$, the angular diameter distance from the observer to the source $D_S$, to the lens $D_L$ and from lens to the source $D_{LS}$ and by the relative position of the image in the image plane $\vect{\xi}$.}
 \label{fig:lensillustration}
\end{figure}

When the distances between the observer, lens, and source are large, the thin lens approximation holds, and two quantities describe the gravitational lens: 
its relative position to the source $\vect{\eta}$ and its two-dimensional projected mass distribution $\Sigma(\vect{\xi})\equiv \int \rho(\vect{\xi},z) dz$, where $\rho(\vect{\xi},z)$ is the density of the lens at a given position ($\vect{\xi}$, $z$) on the image plane.
The source displacement $\vect{\eta}$ and image position $\vect{\xi}$ can be related to the angular coordinates by 
\begin{equation}
\begin{split}
 \vect{\eta}&=D_S \vect{\beta}\,,\\
 \vect{\xi}&=D_L \vect{\theta}\,,
\end{split}
\end{equation}
where $D_S$, $D_L$  and $D_{LS}$ are the angular diameter distances from the observer to the source, to the lens and from the lens to the source, respectively. 
An illustration of the lens configuration is given in Fig.~\ref{fig:lensillustration}.

Here we consider the geometrical optics approximation, which is valid when the GW wavelength is smaller than the characteristic size of the space-time curvature~\citep{takahashi2003wave}. Therefore, we limit ourselves to microlenses that are above $\sim 100 \, \rm M_\odot$. Note, however, that small wave-optics corrections may occur for these scenarios in the low-mass limit (see Sec.~\ref{subsec:waveforms and strains} for discussion on the applicability of this approximation). 

In the geometrical optics limit, the lens focuses the original GW from several paths towards an observer, forming multiple images. 
The image positions and time-delays can be solved from the lens equation
\begin{equation} \label{eq:lensequation}
\begin{split}
 \nabla_{\vect{\theta}} \left[ \frac{1}{2} (\vect{\theta}-\vect{\beta})^2 - \psi(\vect{\theta}) \right] = 0\,,
\end{split}
\end{equation}
where $\nabla_{\vect{\theta}}$ is the two-dimensional nabla operator with respect to $\vect{\theta}$ and $\psi(\vect{\theta})$ is the two-dimensional deflection potential. The latter can be generally solved from the two-dimensional Poisson equation
\begin{equation}
 \nabla^2_{\vect{\theta}} \psi(\vect{\theta}) = 2 \kappa(\vect{\theta})\,,
\end{equation}
where $\kappa = \Sigma(D_L \vect{\theta})/\Sigma_{\rm cr}$, $\Sigma$ is the surface mass density of the lens and $\Sigma_{\rm cr} = (c^2/(4 \pi G)) D_S/(D_L D_{LS})$. 
Solving Eq.~(\ref{eq:lensequation}) yields $N$ different solutions, which give the image positions $\{ \vect{\theta}_j \}$. 

Given the image positions $\{ \vect{\theta}_j \}$, it is possible to retrieve the individual magnifications $\{ \mu_j \}$ and time-delays $\{ t_{d,j} \}$ for each image by direct substitution~\citep{diego2019observational}:
\begin{align}
 t_{d,j}&=t_d(\vect{\theta}_j, \vect{\beta}) = \frac{D_L D_S}{D_{LS}} \frac{1+z_L}{c} \left[ \frac{1}{2} (\vect{\theta}_j-\vect{\beta})^2 - \psi(\vect{\theta}_j) \right]\,,\\
 \mu_j &= \left[ 1/\det \left( \frac{\partial \vect{\beta}}{\partial \vect{\theta}} \right) \right]_{\vect{\theta} = \vect{\theta}_j}\,.
\end{align}

The lensing effect induced on the waveform is expressed by the magnification function
\begin{equation} \label{eq:amplificationfunctiongeometrical}
 F(f;\vect{\lambda}_{\rm lens}) = \sum_j |\mu_j|^{1/2} \exp(2 \pi i f t_{d,j} - i \pi n_j),\quad  f\geq 0 \,,
\end{equation} 
and the lensed waveform is then
\begin{equation}
\begin{split}
 h_{\rm lensed}(f; {\vect{\lambda}}, \vect{\lambda}_{\rm lens}) =&\sum_{j=1}^N \left[ |\mu_j|^{1/2} \exp(2 \pi i f t_{d,j} - i \pi n_j) \right] \\
 &\times h(f; {\vect{\lambda}})\,,
\end{split}
\end{equation}
i.e., a combination of $N$ different signals with magnifications $\mu_j$, relative time-delays $t_{d,j}$ and overall phase shifts $\pi n_j$. Here, $\vect{\lambda}$ and $\vect{\lambda}_{\rm lens}$ indicate the sets of parameters that describe the unlensed GW and the lens, whereas the Morse indices $n_j$ are $n_j=0,1/2,1$ when $\vect{\theta}_j$ is a minimum, saddle and maximum of the time-delay $t_d(\vect{\theta}, \vect{\beta})$, respectively.

\section{\textsc{lensingGW}} 
\label{sec:lensinggw}

In order to find the lensed images of a source, one has to solve the lens equation (Eq. (\ref{eq:lensequation})). That is, a system of two non-linear, algebraic, coupled equations of two variables. In addition, the gravitational potential $\psi$ may not be analytic. No procedure is guaranteed to find a complete set of solutions unless the initial values given to the algorithm are close enough to the actual images~\citep{press1992}. For these reasons, gravitational lensing software packages which provide solutions of the lens equation~\citep{lenstronomy, gravlens} usually rely on a backward procedure: the image plane is tiled into pixels, whose centers (or edges, depending on the package) serve as a dummy input for $\boldsymbol{\theta}$ in the lens equation. 

For a given potential, putative source positions are predicted for each pixel (ray-shooting), and tiles hosting candidate solutions are identified by comparing the source position that they predict to the true one. For example, \textsc{lenstronomy}\footnote{https://github.com/sibirrer/lenstronomy}~\citep{lenstronomy} considers the ones that, when ray-shooted, minimize the distance from the source locally. To achieve better precision, programs like \textsc{lenstronomy} also provide them as seeds to a numerical root finder. 

When stellar-mass lenses are embedded into galaxies or galaxy clusters, microimages are likely to form on top of strongly lensed images and can be separated by as little as $\mu \rm{as}$~\citep{diego2019observational}. Despite circumventing the initial value problem, ray-shooting algorithms that rely on a predetermined pixel size and possibly standard root finders may prove inadequate in such scenarios. Indeed, two or more images may not be identified as separate solutions if they lie in the same pixel.

In order to resolve the microlensed images, the fixed tile must be smaller than the image separations, which are not known a priori. The sky area spanned by strongly lensed images, though, is $\lesssim as^2$~\citep{schneiderBook,collett2015}, and investigating it at such fine resolution is highly time-consuming. Nevertheless, when both galaxies/galaxy clusters and microlenses are involved, a complete solution of the system requires one to identify both strongly lensed and microlensed images.
In addition to that, the overall potential may vary profoundly on short length scales. This is particularly true when many microlenses are involved: standard integrators may stall and oscillate between points corresponding to nearby solutions, without converging. 

Finally, gravitational-wave analysis is performed using signal templates predicted by general relativity (GR). Indeed, the inference of compact source parameters relies on matching gravitational-wave detector strains to a bank of known gravitational waveforms~\citep{Allen_2012}. This methodology allowed LIGO/Virgo to unveil a stellar-mass population of binary black holes and produce a catalog of their properties~\citep{PhysRevX.9.031040}, to perform tests of general relativity~\citep{Abbott_2016, PhysRevLett.123.011102,PhysRevD.100.104036} and to detect a binary neutron star coalescence~\citep{PhysRevLett.119.161101}. However, lensed signals resulting from the superposition of multiple images can differ significantly from unlensed GWs. As a consequence, matched filtering of lensed GWs with unlensed templates could result in missed detections or possibly biased measurements of the binary parameters.

Lensed gravitational waves could also constitute a probe of microlensing. Even when microimages are as close as $\sim \mu \rm{as}$, observable beating patterns may originate in the GW signal if the time delays between the images are above the interferometers' millisecond time resolution~\citep{Lai:2018rto,PhysRevX.9.031040}. Thus, lensed templates could unveil the presence of microlenses and dark matter substructures on the path traveled by the wave~\citep{liao2018anomalies}. 
Therefore, we must be able to predict GW signals from arbitrary lens configurations.

\textsc{lensingGW} has been designed to address these criticalities: on the one hand, it implements a new solver tailored to handle very different scales in the lens potentials; on the other hand, it supplies unlensed and lensed GW signals from arbitrary lensed systems through dedicated modules. Such lensed waveforms can serve as injections in parameter estimation (PE) studies to demonstrate the inference of lensed compact sources' astrophysical parameters through Bayesian analysis. We will demonstrate Bayesian PE of lensed systems solved with \textsc{lensingGW} in a follow-up study~\citep{giuliaforwardcitation}.

Source code of \textsc{lensingGW} and examples are hosted on GitLab, at \url{https://gitlab.com/gpagano/lensinggw}. Documentation is hosted on GitLab pages, at \url{https://gpagano.gitlab.io/lensinggw/}.

\subsection{The lens equation solver}
\label{subsec:lens equation solver}

The numerical solver implemented in \textsc{lensingGW} introduces two novelties: a \textit{two-step procedure} designed to resolve strongly lensed images and microimages simultaneously and a dedicated procedure to handle sources (images) close to the caustics (critical curves), or lines of infinite magnification in the source (image) plane. Moreover, additional features allow us to assess the impact of the components of multi-component lens models and to increase the resolution of the algorithm near the source position. 

\subsubsection{The two-step procedure}
\label{subsubsec:two-step procedure}
The two-step procedure of \textsc{lensingGW} implements two main innovations with respect to previous algorithms:

\begin{enumerate}
    \item It splits the lens system into a \textit{macromodel} (galaxy or galaxy cluster) and a \textit{background} (the remaining lenses);
	\item It selectively zooms on the image plane by iteratively subtiling pixels classified as candidate solutions until the desired precision is met; the number of iterations of such refinement, hence the finest grid resolution, is dynamically adjusted by the algorithm and does not need to be specified in advance.
\end{enumerate}

Images are found in two steps: first, the macromodel is considered and the images formed by it (\textit{macroimages}), are found. The sky region where the search is performed is centered on the source position and its size is specified by the user: images which form outside of it will not be found. The macroimages serve as initial guesses for the image positions of the complete model (macromodel + background). A further inspection is carried on around those, now considering the full lensing system. This significantly reduces the sky area to search out for the images of the complete system and naturally decouples different scales in the lensing potentials.

The rationale is that in realistic lensing scenarios where microlenses are embedded in galaxies or galaxy clusters, it is possible to identify a dominant deflector (the galaxy/galaxy cluster) and the remaining lenses (the microlenses) represent a small perturbation to its potential. Thus, their effect is that of splitting or shifting the macroimages by a small amount with respect to the macroimage separations. Their locations can then guide the inspection of the complete model.
Image separations may still be large if that is not the case, but one can nevertheless set a larger window for the complete model and cover all the relevant areas. Therefore, the definition of macromodel is arbitrary: it only establishes which deflector is considered first. For instance, the macromodel can be composed of several lensing components, such as a bulge, a disk, and a halo. 

\begin{figure}
\centering
 \includegraphics[width=\columnwidth]{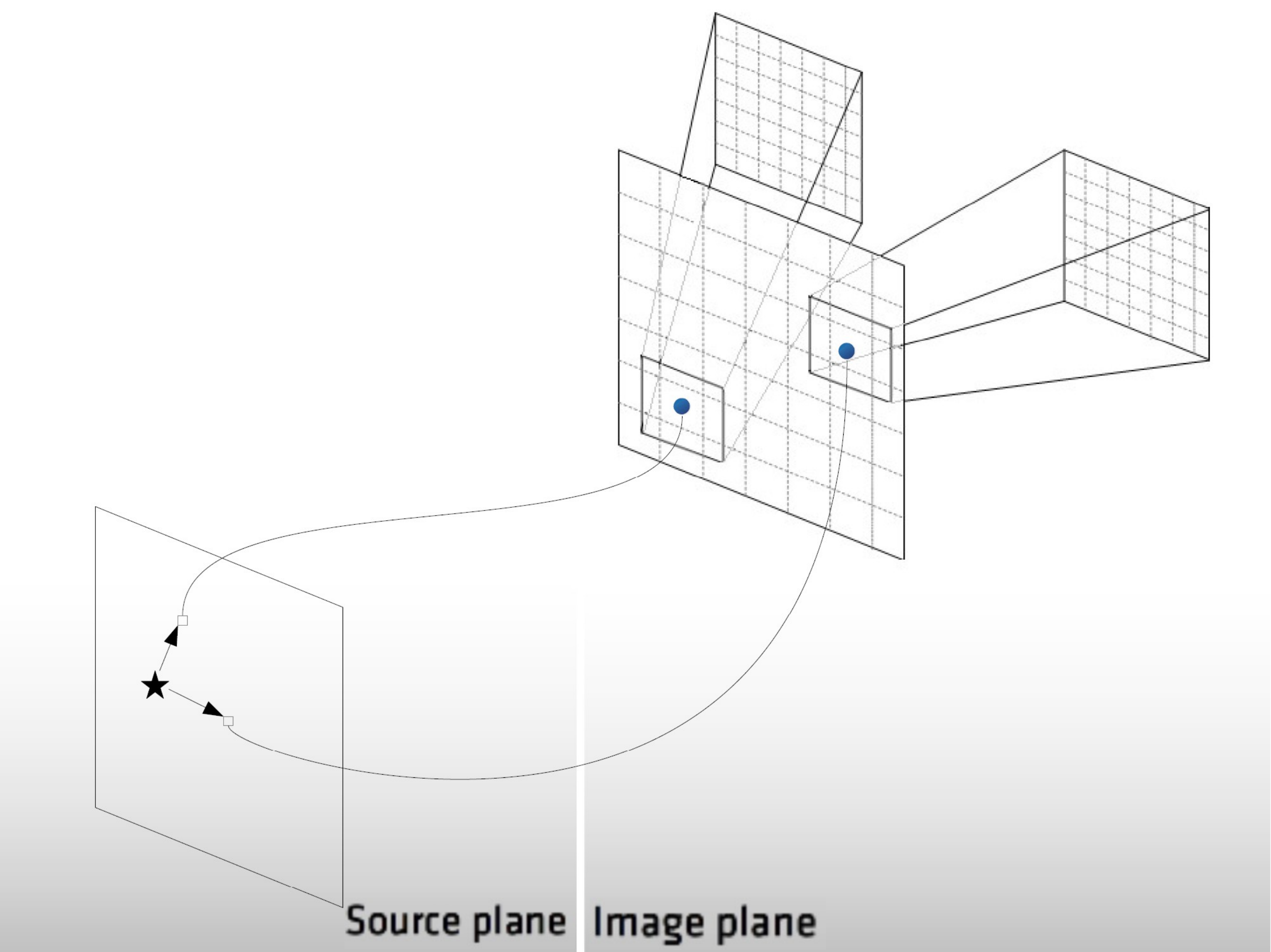}
 \caption{\textit{Iterative procedure of \textsc{lensingGW}}. The image plane is ray-shooted to the source plane via the lens equation. Pixels whose projected distances from the source (the star) are local minima are iterated over through adaptive grids. The process stops when no more minima are found or a minimum pixel size is reached.}
 \label{fig:mysolver}
\end{figure}

At both steps, the search windows are tiled and pixels classified as candidate solutions are found through ray-rooting, as per what is implemented in \textsc{lenstronomy}. This routine is justified by the fact that true solutions must exactly reproduce the source position when ray-shooted. Thus, approximate solutions should minimize their ray-shooted distance from the actual source locally. Pixels that already meet the desired precision are stored as images, while all the other tiles hosting a candidate solution are iterated over: they constitute the centers of the next grids, whose sides are twice the size of the original pixel, to avoid missing solutions at the boundaries. The tiling and the iteration are repeated on the new grids until no more candidate regions are found or the safety threshold pixel size of $10^{-25}\, \rm{rad} \, (\sim 10^{-20}\,\rm{as})$ is reached. This is a conservative choice motivated by the fact that typical microimages separations are orders of magnitude above such threshold. We recall here that recent studies have found them to form as close as $\mu \rm{as}$~\citep{diego2019observational}. Thus, if microimages have not been identified in the zoomed region at such resolution, the iteration is likely to be driven by numerical artifacts and can be interrupted without loss of solutions. This process prevents the algorithm from getting stuck between nearby images and guarantees that each interesting area is checked at a finer resolution. The procedure is exemplified in Fig.~\ref{fig:mysolver}.

The advantage of our approach compared to methods based on fixed tiles or a predetermined number of subtiles is that it allows the identification of both strongly lensed and microlensed images on a broad sky region while retaining fast performance and assuming no \textit{a priori} knowledge on the image separations. Indeed, the zoom is performed on each area on a case-by-case basis: regions, where no images are present will not be zoomed at all. Strongly lensed images will require a moderate amount of iterations to meet the desired precision. At the same time, more layers of grid refinement will be employed where microimages are present, allowing us to resolve all the images. The tile and grid extensions are adjusted dynamically in each area, and at each iteration; therefore, the refinement is focused only where needed.

On the other hand, as previously mentioned, methods that rely on fixed tiles or subtiles can identify arbitrary images' separations only when the whole sky region is tiled at the finest needed resolution, which requires the scale of such separations to be known in advance. \textsc{lensingGW} can overcome these limitations.

We demonstrate the performance of \textsc{lensingGW} with respect to the fixed tile approach on a system which produces both strongly lensed images and microimages. We consider an elliptical galaxy at the origin of the image plane with mass $M_G = 10^{12}\rm{M_\odot}$, ellipticity $\epsilon=0.1$ and core radius $R_c = 500\,pc$ at redshift $z_L=0.5$. The galaxy potential is
\begin{equation}
\psi(\vect{\theta}) = \theta_E \sqrt{\theta_c^2+(1-\epsilon)\theta_x^2+(1+\epsilon)\theta_y^2} \, ,
\end{equation}
where $\theta_c = R_c/D_L$, $\theta_E = \sqrt{(4GM_G/c^2)D_{LS}/D_LD_S}$ is the Einstein radius and $\vect{\theta}=(\theta_x,\theta_y)$ is the two-dimensional coordinate in the image plane. A point source placed inside the diamond shaped caustic at redshift $z_S=2$ and sky position $(ra, dec) = (0.05,0)\cdot\theta_E\,\rm{rad}$ produces five strongly lensed images, whose time delays span from $\sim$ days to $\sim$ months and with total magnification $\mu=\sum_{i\in{\mathrm{macroimages}}}|\mu_i| = 21.3$. To produce microlensing in one region of the macromodel, we inject a background of $\sim600$ point mass microlenses with masses $m_i\in[100,200]\rm{M_\odot}$ in an area of $\sim$ 14\rm{mas} x 14\rm{mas} around the position of the most magnified macroimage. The microlenses are uniformly distributed in mass and position until the target density of $12\rm{M_\odot/pc^2}$ is reached\footnote{This is the same surface mass density as the one adopted in~\citep{diego2019observational} for lensed extragalactic gravitational waves at high macromodel magnifications.}.

\begin{figure*}
    \includegraphics[width=\textwidth]{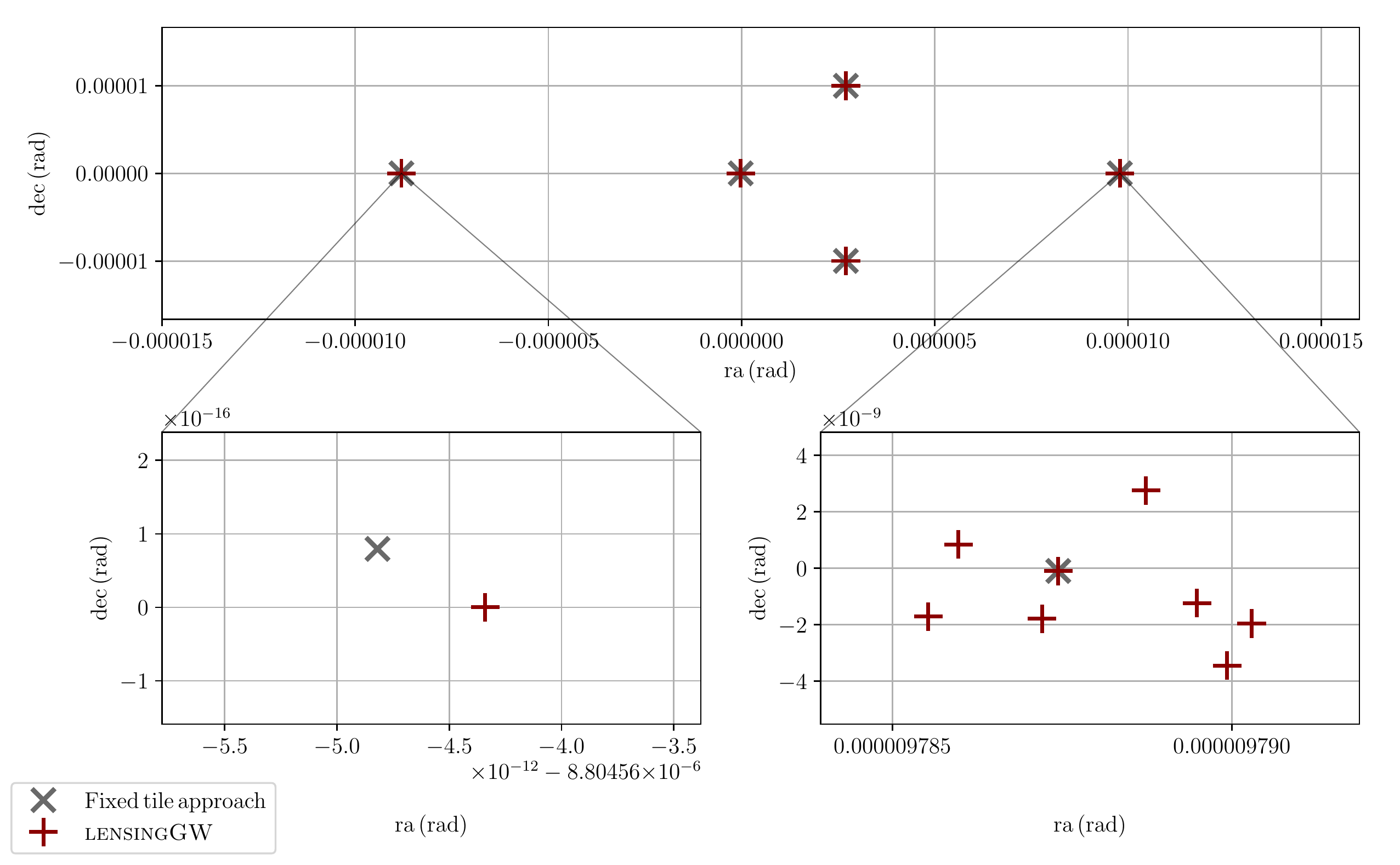}
\caption{\textit{Comparison between \textsc{lensingGW}'s lens equation solver and a fixed tile algorithm}. Right ascension (ra) and declination (dec) of the images recovered by \textsc{lensingGW} (red crosses) and by a fixed tile routine (grey \romannumeral 10) for an elliptical galaxy plus a background of $\sim 600$ microlenses injected around the most magnified macroimage. The strongly lensed images (\textit{top panel}) are correctly recovered by both the algorithms. However, the zoom on the microlensed image (\textit{bottom right}) shows that \textsc{lensingGW} is able to recover $8$ microimages, while the fixed tile approach identifies them as a single one. The solution of the system with \textsc{lensingGW} required $\sim 19\rm{s}$ on a standard machine and $\sim 12\rm{h}$ with the fixed tile method at the lowest tile resolution below the microimage separations. The zoom on a strongly lensed image where no microlens background is present (\textit{bottom left}) is also shown for completeness. Coordinates are relative to the galaxy center.}
\label{fig:lensinggw_vs_fixed}
\end{figure*}

We present the images found by
\textsc{lensingGW} and by the fixed tile ray-shooting in Fig.~\ref{fig:lensinggw_vs_fixed}.
Both programs inspected a sky region of $\sim$ 4.7\rm{as} x 4.7\rm{as} centered on the source position.
\textsc{lensingGW} required $\sim0.044\%$ of the runtime employed by the fixed tile algorithm, inspected $\sim 1/4$ of the pixels, and recovered the images denoted by red crosses in the figure. The pixel size of the fixed tile approach was set to the lowest resolution below the minimum microimage separation. However, it was still too large to resolve the shape of the potential between the microimages and identify the individual solutions: this latter method identified only the macroimages (grey \romannumeral 10). Thus, one would need to try smaller and smaller pixels until the microimages are recovered by the latter method, resulting in a much higher runtime.

The solver can be applied to arbitrary lensing potentials and mass distributions. It accepts lens profiles with the same structure as the ones defined in \textsc{lenstronomy}~\citep{lenstronomy} and is compatible with every lensing potential already implemented therein. The lensing framework (the combination of the lenses into a lens model, the ray-shooting, the selection of pixels classified as candidate solutions, and the computation of caustics and critical curves) derive from \textsc{lenstronomy}.

\subsubsection{Near-caustic sources}\label{subsubsec:optimisation}

When the source approaches the caustics, or curves of infinite magnification in the source plane, the grid refinement may identify an increasingly large number of pixels that could be candidate solutions. In these cases, spurious approximate minima are present at a given iteration. However, by default, the solver iterates on each of them.

The user can then aid the solver to convergence by introducing a cut on the candidate solutions. This feature relies on the fact that regions hosting solutions of the lens equation are expected to improve their ray-shooted distance to the actual source at every iteration. Thus, the user can indicate an improvement criterion to select pixels classified as candidate solutions. In particular, they can set:

\begin{enumerate}
	\item A minimum ray-shooted distance threshold $d^s_0$ to select among pixels classified as candidate solutions in the first grid. This allows for further screening of the interesting regions before the iteration. It should be used when the number of good pixels identified by the algorithm is extremely high already in the first grids;
	\item  How much this distance should improve between subsequent iterations for the candidate solution to be still iterated over. I.e., a multiplicative factor $\delta$, $0<\delta\leq1$ which corrects $d^s_0$ as $d^s=d^s_0\cdot\delta^{n}$ at the $n^{th}$ iteration. It allows us to select among unreasonable numbers of pixels during the procedure.
\end{enumerate}

Combining these functionalities has proven to overcome non-converging cases, allowing \textsc{lensingGW} to retrieve the correct solutions. We illustrate this through the elliptical galaxy of the previous example: we remove the background microlenses and consider a smaller galaxy mass of $M_G=10^{10}\rm{M_\odot}$ to reduce the area of influence of the galaxy and investigate a smaller window. We move the source towards the caustics until the solver without cut does not converge anymore. We then apply the cut with $d_0^s=10^{-7}\,\rm{rad}$ and improvement factor $\delta=0.1$, which leads to the correct identification of all the five macroimages. We record the number of pixels selected by the algorithm as the iteration proceeds and contrast it to the tiles selected when no cut is applied. The comparison is shown in Fig.~\ref{fig:caustic_demonstration}:
\begin{figure}
    \includegraphics[width=\columnwidth]{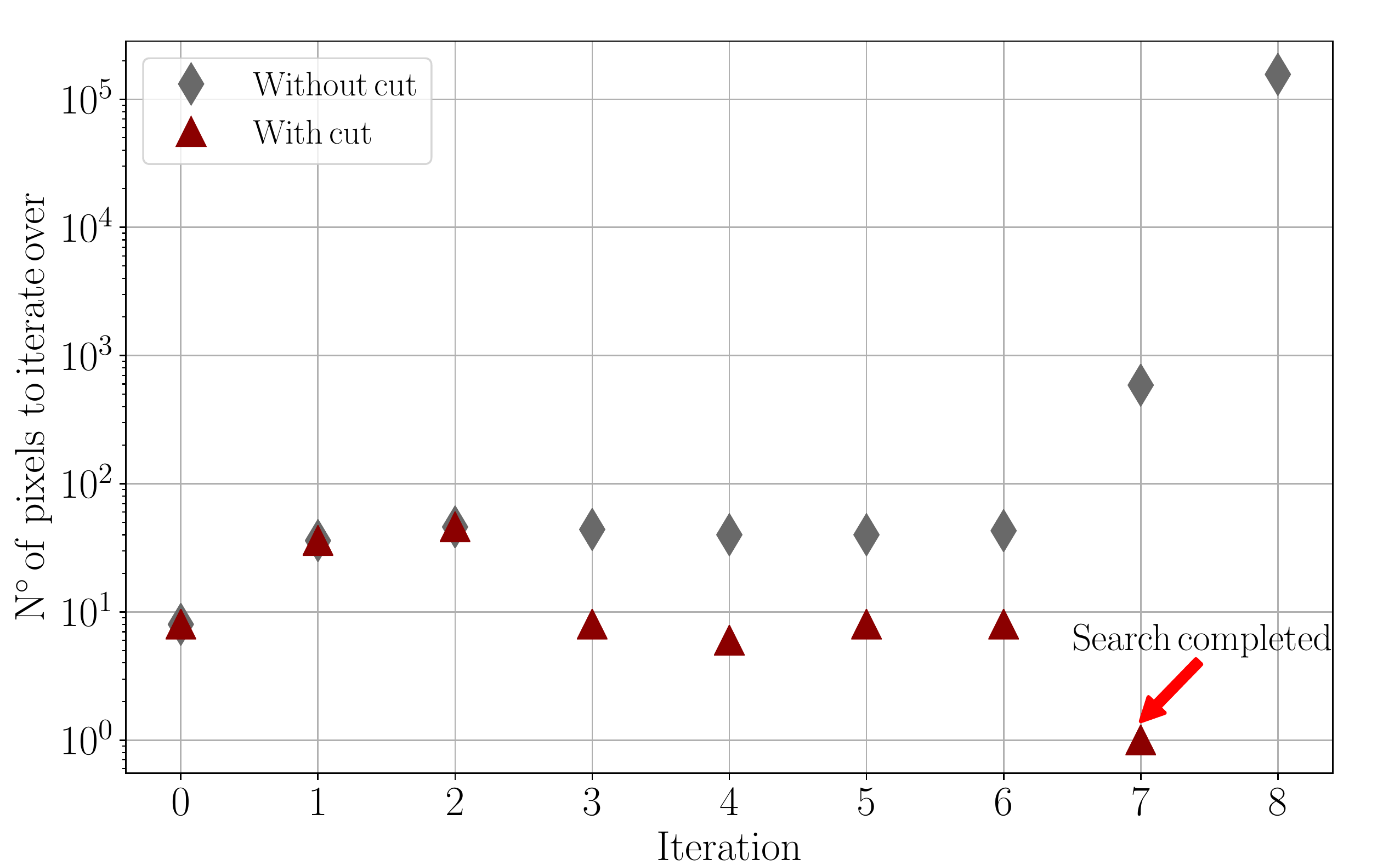}
\caption{\textit{Performance of the improvement criterion for pixel selection over the default selection method for near-caustic sources}. The number of candidate solutions selected for the zoom at each iteration by the default solving algorithm (grey diamonds) and by the improvement criterion (red triangles) for the elliptical galaxy of~Fig.\ref{fig:lensinggw_vs_fixed}, when the source approaches the caustics. The improvement criterion controls the number of tiles to iterate over and recovers the five macroimages in seven iterations, whereas the standard selection routine identifies an ever-increasing number of pixels, leading to more than $10^5$ tiles selected for the zoom at the $8^{\rm{th}}$ iteration.}
\label{fig:caustic_demonstration}
\end{figure}
the improvement criterion keeps the number of candidate solutions under control and leads the solver to convergence in a reasonable number of iterations. On the contrary, the standard setup leads to an increasingly large number of selected tiles, with more than $10^5$ pixels possibly hosting candidate solutions at the $8^{\rm{th}}$ iteration.

By default, \textsc{lensingGW} does not apply the cut, as it discards potential solutions: when no spurious minima are identified, its use could result in missed images. Therefore, the cut is recommended only when the number of candidate pixels increases dramatically, preventing the solver from converging.

\subsubsection{Additional features}\label{subsubsec:features}

The user can inspect the macromodel separately from the microlenses by enabling the \texttt{OnlyMacro} flag in the solver settings. This is important, for instance, when assessing the impact of the background on a galaxy/galaxy cluster. Within \textsc{lensingGW}, the study of the macromodel without (with) background can be performed by enabling (disabling) this functionality. For example, if the elliptical galaxy of Fig.~\ref{fig:lensinggw_vs_fixed} is marked as macromodel and \texttt{OnlyMacro} is active, \textsc{lensingGW} identifies the five strongly lensed images displayed in the top panel of the figure. When the flag is disabled, instead, the elliptical galaxy and microlens background are considered together, and the microimages in the bottom right panel are recovered in addition to the strongly lensed ones.

In the same manner, the user can flag certain components of a lens model (e.g., a bulge and a disk) as a part of the macromodel, and the remainder as part of the background (such as a halo). The user can then use the \texttt{OnlyMacro} option to study the model with and without the background. We recall here that this is possible because the definition of macromodel is arbitrary and only indicates which lens models are considered for the first step of the solving algorithm.

Depending on the lens position, the images can also form close to the position where the image in the absence of lenses would form (the unlensed image position).
The user can require a further zoom close to the unlensed image position through the \texttt{NearSource} flag. Should they wish to span a wider area around the lens while retaining details close to the unlensed image, this functionality allows to specify two windows with a different number of pixels for the first step. Thus, one window can be used to investigate the larger region and cover the whole lens, while the other one can be used to span a much smaller area around the unlensed image position, increasing the resolution where it may be needed. 

\subsection{GW signals}\label{subsec:waveforms and strains}

Once we have found the number of images and the relative time delays of a lensing configuration, we typically wish to do further parameter estimation on the signal. Parameter estimation studies of compact binary coalescences (CBCs) are performed through templates of gravitational signals predicted by general relativity (GR) and their projection into the detectors, or strain. Templates are needed to assess the detectability of the signals and the characteristics of the source.
If unlensed gravitational-wave strains are used in filtering microlensed gravitational-wave data, the search method which allows us to identify a signal in the detector data (\textit{matched filtering}~\citep{Allen_2012}) becomes sub-optimal. 

\begin{figure}
    \centering
    \includegraphics[width=\columnwidth]{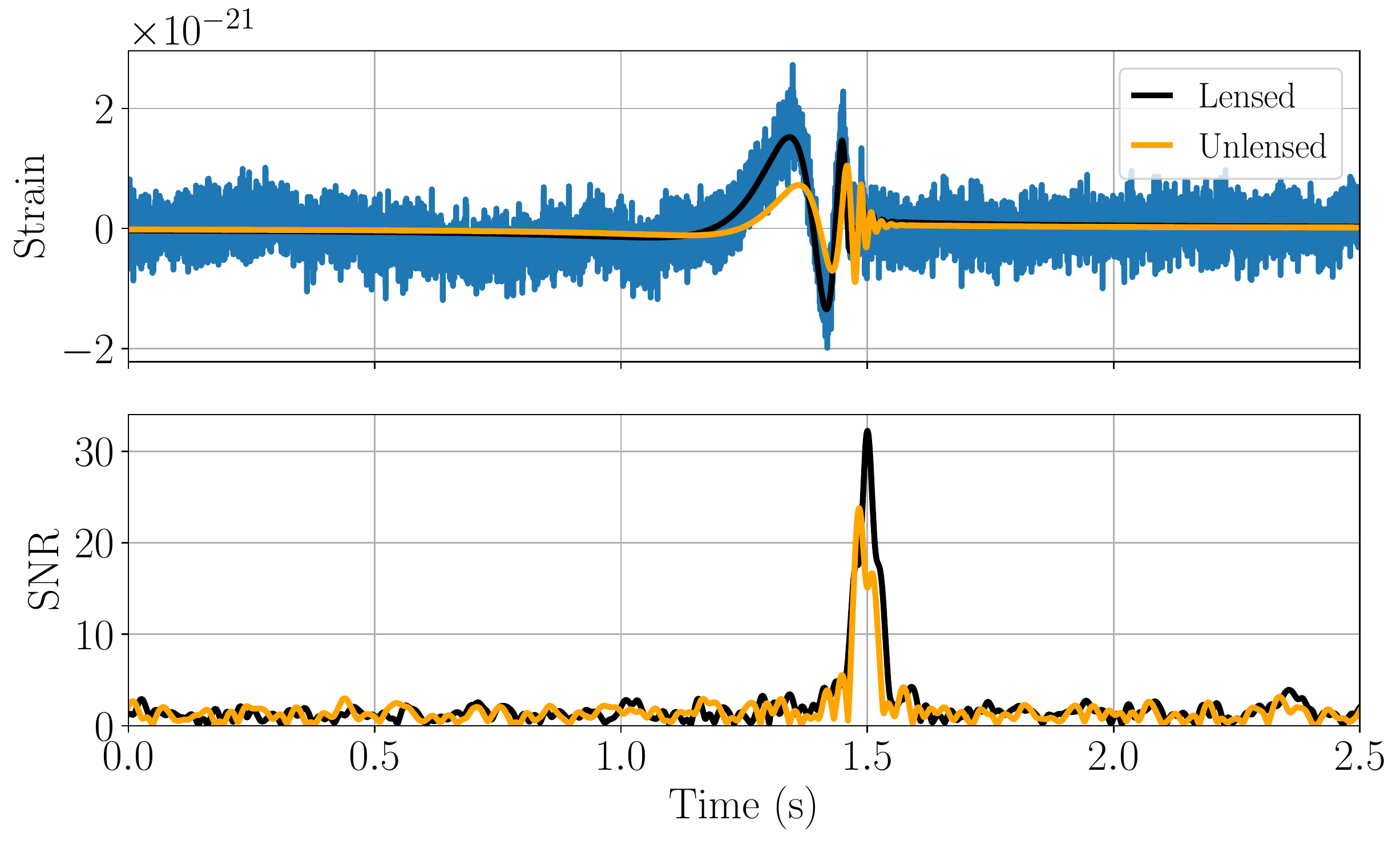}
    \caption{\emph{Illustration of matched filtering with lensed and unlensed waveforms.} 
    \emph{Top panel:} An example microlensed gravitational-wave strain (black) and corresponding detector data with noise (blue). The unlensed strain  is also shown for comparison (orange). 
    \emph{Bottom panel:} Matched filter signal-to-noise ratio of the detector data using the microlensed waveform (black) and the unlensed waveform (orange). The microlensed gravitational-wave strain matches the data better and thus returns a higher matched signal-to-noise ratio. 
    Here, a high signal-to-noise ratio example is considered for illustration and both signals are normalized so that their optimal signal-to-noise ratios are the same.
    }
    \label{fig:snrtimeseries}
\end{figure}

As an illustrative example, we show the filtering process for a microlensed gravitational waveform within a LIGO detector at design sensitivity. We simulate a gravitational wave from a binary black hole of masses $m_1 = m_2 = 150\rm M_\odot$, microlensed into two images of comparable magnifications and relative time delay of $10 \rm ms$. We filter the data using a lensed and an unlensed strain, finding that the unlensed waveform returns a lower matched-filter signal-to-noise ratio (Fig.~\ref{fig:snrtimeseries}).
Hence, it is fundamental that software packages for lensing of CBCs come with an infrastructure for gravitational waves that predicts GWs from arbitrary lensed systems. \textsc{lensingGW} provides such infrastructure through specific modules that compute unlensed templates and strains through \textsc{LALSimulation}~\citep{lalsuite} and their lensed counterparts in ground-based detectors. Thus, any waveform development in the LIGO Scientific Collaboration Algorithm Library Suite \textsc{LALSuite}~\citep{lalsuite} will benefit \textsc{lensingGW}. 

\begin{figure}
    \centering
    \includegraphics[width=1\columnwidth]{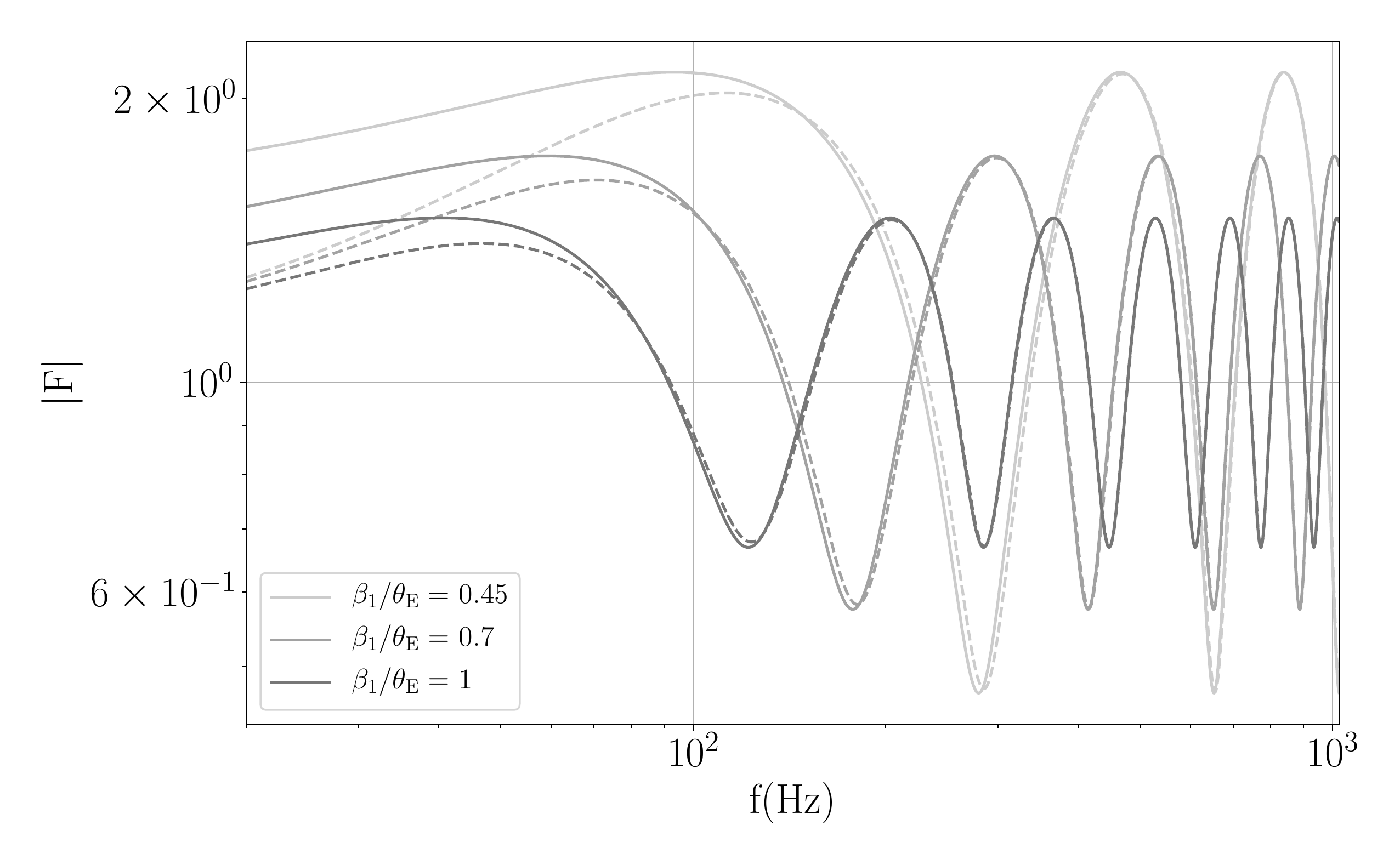}

\caption{\textit{Comparison between the geometrical optics amplification and the wave optics amplification for an isolated point mass}. We contrast the geometrical optics approximation (full line) to the wave optics magnification (dashed line) for an isolated point lens of mass $M_L=100M_\odot$ at the origin of the image plane and source position $\vect{\beta}=(0,\beta_1) \, \rm{rad}$. The error in the geometrical optics approximation rapidly reduces for frequencies $f\gtrsim 100 \rm Hz$ and systems with higher source displacements.
}
\label{fig:geom_vs_wave}
\end{figure}

Magnifications are currently computed within the geometrical optics approximation of Eq.~(\ref{eq:amplificationfunctiongeometrical}). The approximation is valid for galaxies and galaxy clusters at LIGO/Virgo frequencies, but may not be valid in general for small isolated lenses when wave optics effects become important. In Fig.~\ref{fig:geom_vs_wave}, we show the geometrical optics and wave optics magnifications for an isolated point lens of mass $M_L = 100M_\odot$. The difference between the two approaches rapidly reduces in the most sensitive frequency band of the interferometers ($f \gtrsim 100 \rm Hz$) and for higher source displacements. However, one should practice caution with smaller isolated microlenses and source-point lens separations. Therefore, here we focus on lenses with $M_L \gtrsim 100 \, \rm{M_\odot}$ and source-point lens displacements $\gtrsim 0.45\cdot\theta_E$. Future extensions will cover wave optics, but we will not cover it here.

\section{Validation}
\label{sec:validation}

We test the solver on two scenarios proposed in the literature: a microlens embedded in a host galaxy at high magnifications and two equal mass point lenses. The two models aim to show that \textsc{lensingGW} can correctly recover the lensed images when dealing with potentials that involve different scales and with comparable masses, hence demonstrating that it is a valid tool for generic lensing configurations.

\subsection{Diego et al.~\citep{diego2019observational}} \label{sec:diego et al}

Diego et al.~\citep{diego2019observational} propose, among many other models, the case of a point mass microlens of $M_L = 100M_\odot$ embedded in a galaxy/galaxy cluster at redshift $z=0.5$.
The galaxy is described by a potential with both convergence and external shear and with total magnification $\mu=30$. We consider the configuration in which a source is placed at redshift $z=2$ and angular position $\beta/\theta_E = 0.05$, where $\theta_E = \sqrt{(4GM_L/c^2)D_{LS}/D_LD_S}$ is the Einstein radius of the microlens. This case leads, in the positive parity side of the galaxy (i.e., where the images are minima or maxima of the time delay), to four microimages, two of which lie within the critical curves and two outside them (see Fig. 5 of~\citep{diego2019observational}).

\begin{figure}
\centering
 \includegraphics[width=1\columnwidth]{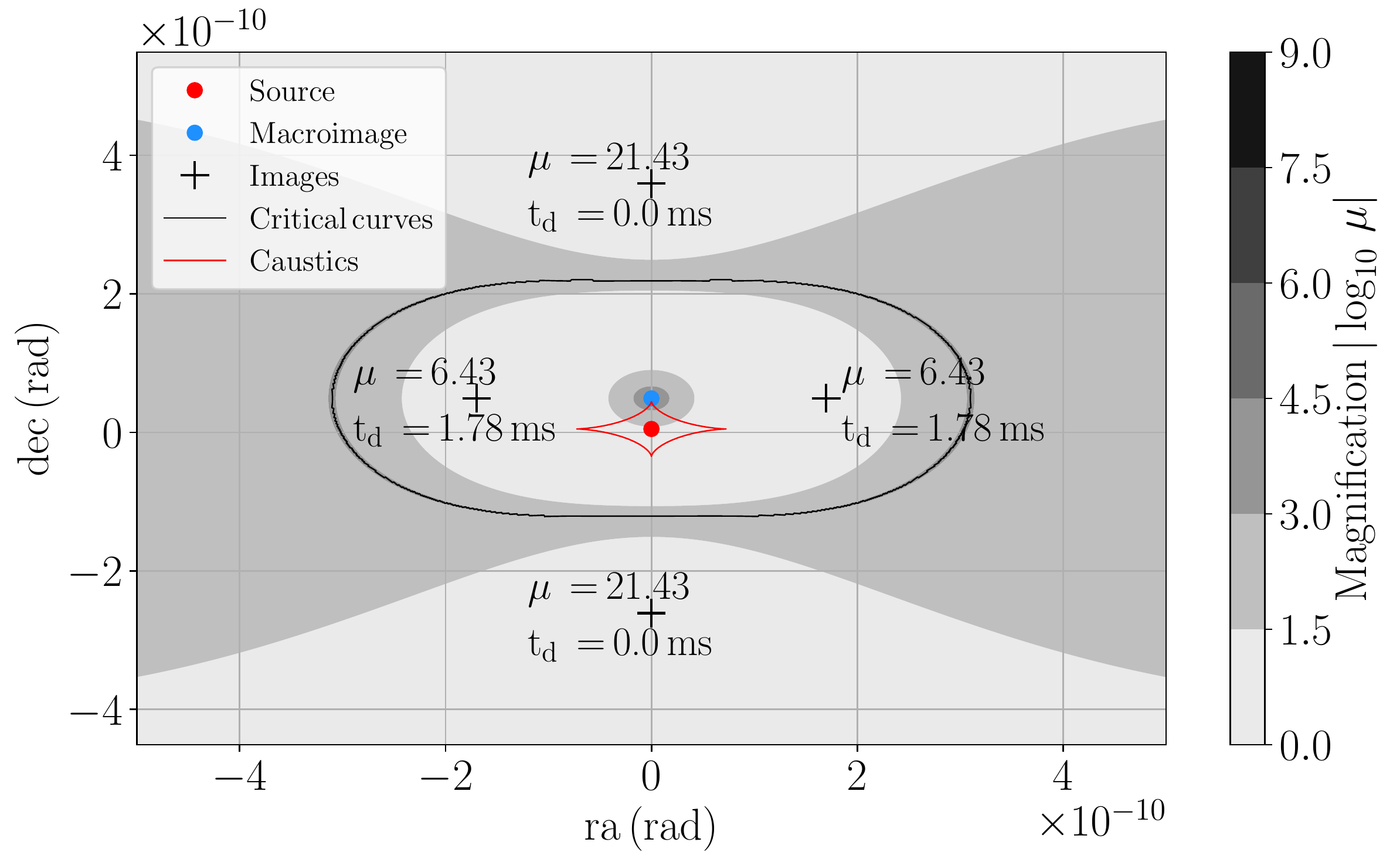}
 \caption{\textit{Microimages recovered by \textsc{lensingGW} for the galaxy plus point lens system of Diego et al.~\citep{diego2019observational}}: right ascension (ra) and declination (dec) relative to the galaxy center. The off-axis source (red dot) is lensed by the galaxy and a shifted image forms (blue dot). When a $100M_\odot$ point lens is added at the position of the macroimage, it splits it into four microimages (black crosses). Two of them form inside the critical curves (black line) and two outside of it with magnifications $\mu$ and normalized time-delays $t_d$.
 The caustics (red line) and a color map of the magnification in the ra-dec plane are shown for completeness. The system demonstrates the applicability of \textsc{lensingGW} to microlenses embedded in galaxies/galaxy clusters.
 }
 \label{fig:diego_et_al}
\end{figure}

We set the galaxy to be the macromodel and define the point lens as the background, which we place at the macroimage position. The images recovered by \textsc{lensingGW} for this scenario are illustrated in Fig.~\ref{fig:diego_et_al}.

The image geometry is correctly recovered with respect to the critical curves. Magnifications, time delays, and image separations are comparable to what was found in Ref.~\citep{diego2019observational}. The exact source and point lens positions are not specified in Diego et al. The numerical values of the recovered quantities are not identical for this reason.

\subsection{Schneider, Wei{\ss}~\citep{schneiderWeiss}}

We now consider the case of two point lenses of equal masses. Schneider and Wei{\ss}~\citep{schneiderWeiss} have extensively studied this scenario: the authors calculate the number of images and their approximate positions for a variety of source positions and lens displacements. Images are identified by specifying the quadrant of the image plane they form in and whether they are inside the critical curves or not.

\begin{figure}
\centering
 \includegraphics[width=1.\columnwidth]{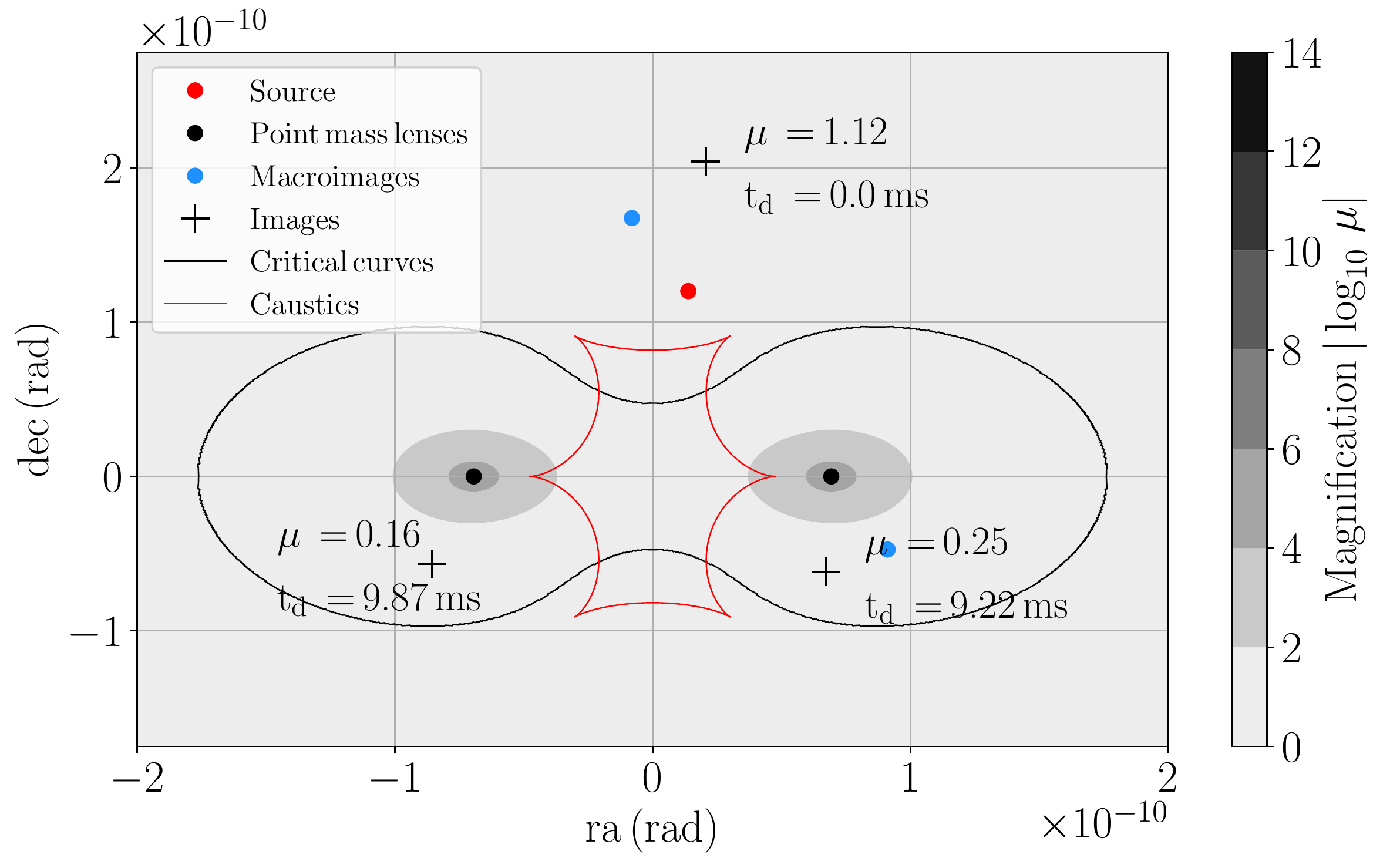}
 \caption{\textit{Images recovered by \textsc{lensingGW} for a binary point lens}: right ascension (ra) and declination (dec) relative to the binary center of mass. The source (red dot) is lensed by two point masses (black dots) of $M_1 = M_2 = 100M_\odot$. When only one deflector is considered, two macroimages (blue dots) form. When the second lens is added, three total images (black crosses) form: their positions with respect to the critical curves and to the origin are consistent with the predictions of Schneider and Wei{\ss}. Magnifications ($\mu$), normalized time delays ($t_d$), caustics (red line) and a color map of the magnification in the ra-dec plane are shown for completeness. The system confirms the applicability of \textsc{lensingGW}'s algorithm to comparable-mass potentials.}
 \label{fig:schneider_weiss}
\end{figure}

Seven configurations from Ref.~\citep{schneiderWeiss} have been tested: in Fig.~\ref{fig:schneider_weiss} we show the off-axis case corresponding to $\vect{\beta}/\theta_E = (0.1,0.5\sqrt{3})$ and lens positions $\vect{x/\theta_E} =(\pm0.5,0)$, where $\theta_E$ is the Einstein radius given by the total mass $M_{L1}+M_{L2}=200M_\odot$. Source and lenses are placed at redshift $z=2$ and $z=0.5$, respectively. Here, we consider the point lens closest to the source as macromodel and the farthest as background. The same result is obtained when both lenses are indicated as macromodel simultaneously, and the \texttt{OnlyMacro} option is enabled (see Sec.~\ref{subsubsec:features}). Indeed, we recall that the macromodel can be formed by multiple lenses and its definition is arbitrary.

For this scenario, the authors predict three images: one outside the critical curve in the first quadrant and two inside the critical curve, in the third and fourth quadrant. The geometry illustrated in Fig.~\ref{fig:schneider_weiss} shows that \textsc{lensingGW} is able to recover the correct solutions.

\section{Applications}
\label{sec:applications}

In this section, we demonstrate possible applications of \textsc{lensingGW} to scenarios of astrophysical interest. First, we use it to perform a parameter space investigation of the source parameters and a systematic detectability study of the resulting lensed GWs. We then apply it to investigate the effects of strongly lensed images' properties on microlensing, in a lensing system with hundreds of microlenses embedded within a galaxy. In what follows, we consider ground-based gravitational-wave detectors at design sensitivity.

We recall here that the inner product between two signals $a$ and $b$ is defined as~\citep{cutlerFlanagan}:
\begin{equation}
 \langle a, b \rangle = 4\Re\int_{f_{\rm min}}^{f_{\rm max}}\frac{a(f)b^*(f)}{S_n(f)} df\,,
\end{equation}
where $a(f)$ and $b(f)$ are the Fourier transforms of $a(t)$ and $b(t)$, $^*$ denotes complex conjugation, $S_n(f)$ is the one-sided power spectral density (PSD) of the instrument and $f_{\rm min}$ and $f_{\rm max}$ define the relevant frequency band.

The agreement between two GW signals $h_1$ and $h_2$ is commonly quantified as the noise-weighted inner product of the normalized waveforms, maximized over the time and phase of coalescence:
\begin{equation}
 M(h_1,h_2) =\max_{t_c,\phi_c}\langle\hat{h}_1,\hat{h}_2 \rangle\,,
\end{equation}
where $\hat{h} = h/\sqrt{\langle h,h \rangle}$ is the normalized waveform. When two signals have identical phasing, this statistics, or match, equals one: their distinguishability is then quantified through the mismatch: $\mathcal{M}(h_1,h_2)=1-M(h_1,h_2)$.

\subsection{Parameter space investigation and detectability study}
\label{subsec:diego mismatch}

We show the impact of the component masses of the compact source on the distinguishability of lensed signals for the lens model of Section~\ref{sec:diego et al}. We simulate a compact source\footnote{The unlensed signal is produced by a non spinning binary with sky position $(ra,dec) =(0,0.05\cdot \theta_E) \, \rm{rad}$, redshift $z=2$ and inclination $\iota= 2.6 \, \rm{rad}$.} in the Hanford detector and vary the black hole masses in the $\mathcal{M}_c-q$ plane, where $\mathcal{M}_c = (m_1m_2)^{3/5}/(m_1+m_2)^{1/5}$ is the chirp mass and $q =m_2/m_1\leq1$ is the mass ratio of the binary. For each configuration, we predict the source images  and model the lensed and unlensed signals in the LIGO detector through \textsc{lensingGW}. We then compute their mismatch: signals with larger mismatches are more likely to be distinguishable as lensed.

\begin{figure}
 \centering
 \includegraphics[width=1.\columnwidth]{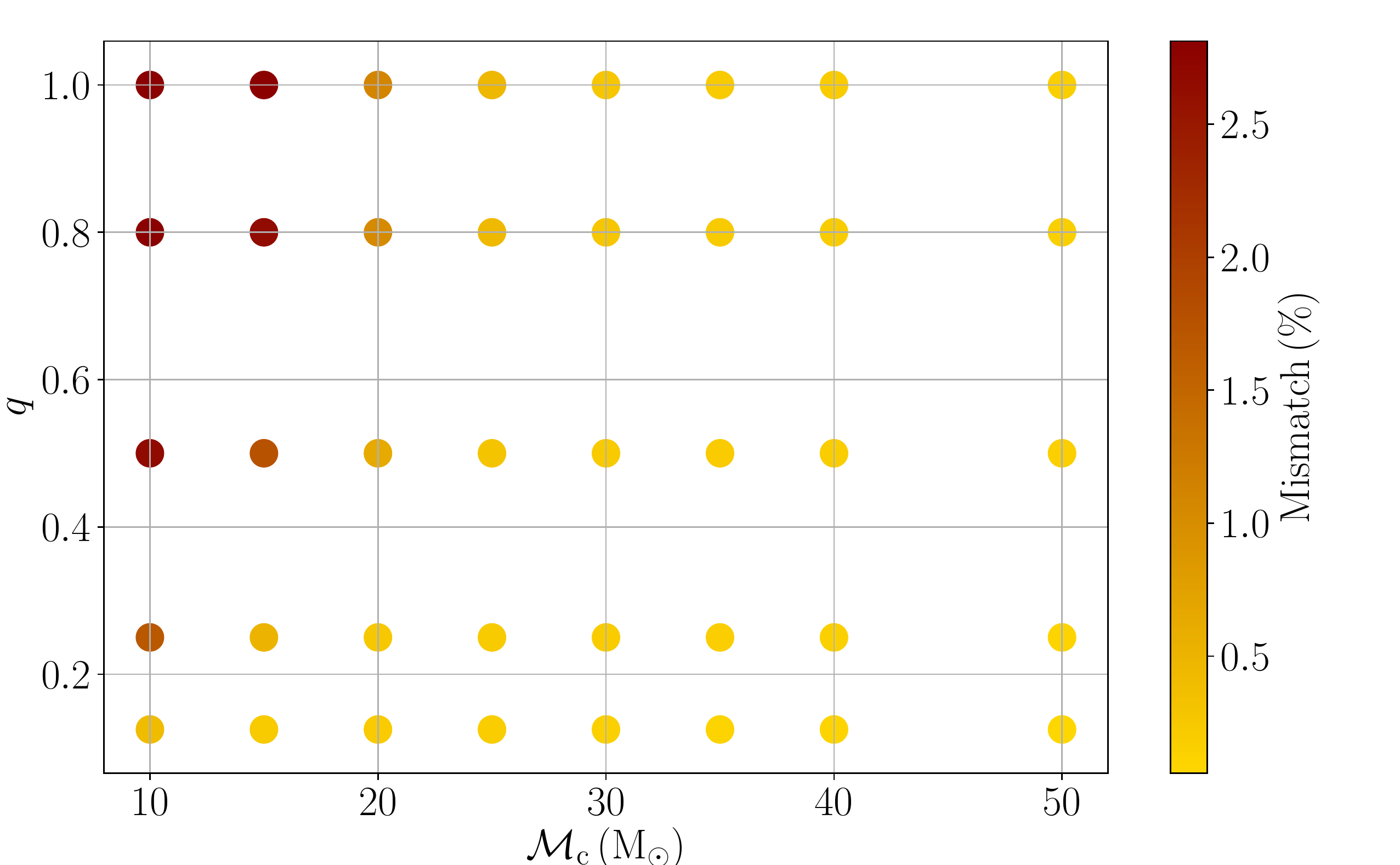}
 \caption{\textit{Source parameter space investigation with \textsc{lensingGW}}: mismatch study of a non spinning compact binary. Each dot represents a pair of component masses in  the $\mathcal{M}_c-q$ plane, lensed by the galaxy plus point lens model presented by Diego et al.~\citep{diego2019observational}. All other source parameters are fixed. The point lens mass is $M_L = 100 \, \rm{M_\odot}$ and is placed at the position of the macroimage formed by the galaxy. For each configuration, the mismatch between the lensed and unlensed GWs is displayed in a color-coded map, with darker colors denoting higher mismatches (see the color bar on the right).
 Signals with more significant mismatches are more likely to be distinguishable as lensed. 
 } 
 \label{fig:mismatch}
\end{figure}

As shown in Fig.~\ref{fig:mismatch}, higher mass ratios and lower chirp masses exhibit higher mismatches. This is expected since such signals span a broader frequency band and the dephasing due to the superposition of the images can accumulate over a longer interval. However, the computation of the microimages and the study of the lensed and unlensed strains is essential to assess how the mismatch varies with the source parameters or the lens model. Thus, it is crucial to develop software like \textsc{lensingGW}, which incorporates both the physics of lensing and GW signals while retaining fast performance. 

\subsection{Impact of strongly lensed images on microlensing}
\label{subsec:elliptical macromodel}
We demonstrate how \textsc{lensingGW} can be used to study the impact of a macromodel on a microlens background. We investigate the strain produced by the microimages induced on top of a strongly lensed image by a fixed microlens background, for different magnifications and parities of the strongly lensed image.

We consider the elliptical galaxy plus background of Subsec.~\ref{subsec:lens equation solver} with galaxy mass $M_G=10^{12}\rm{M_\odot}$ and investigate three source positions: $(0.05,0) \cdot\theta_E\, \rm{rad}$, $(0.15,0)\cdot\theta_E\, \rm{rad}$ and $(0.18,0) \cdot\theta_E\,\rm{rad}$. In this scenario, the larger the source position, the closer the source is to the caustic. In this way, we can inspect the change in the results when the magnification of each macroimage is enhanced, while retaining the total number and geometry of the strongly lensed images. 

For each scenario, we consider both the most magnified image in the positive parity side of the galaxy (i.e., where the image is a minimum or a maximum of the time delay) and the one in the negative parity side (where the image is a saddle point). The two sides are expected to behave differently as the region of low magnification is larger than the region of high magnification, resulting in a higher probability of a GW being demagnified by
the microlens than magnified by it. See Ref.~\citep{diego2017dark,Diego:2018fzr,diego2019observational} for a more in-depth discussion.

\begin{figure*}
    \begin{subfigure}
    \centering
    \includegraphics[width=1.\textwidth]{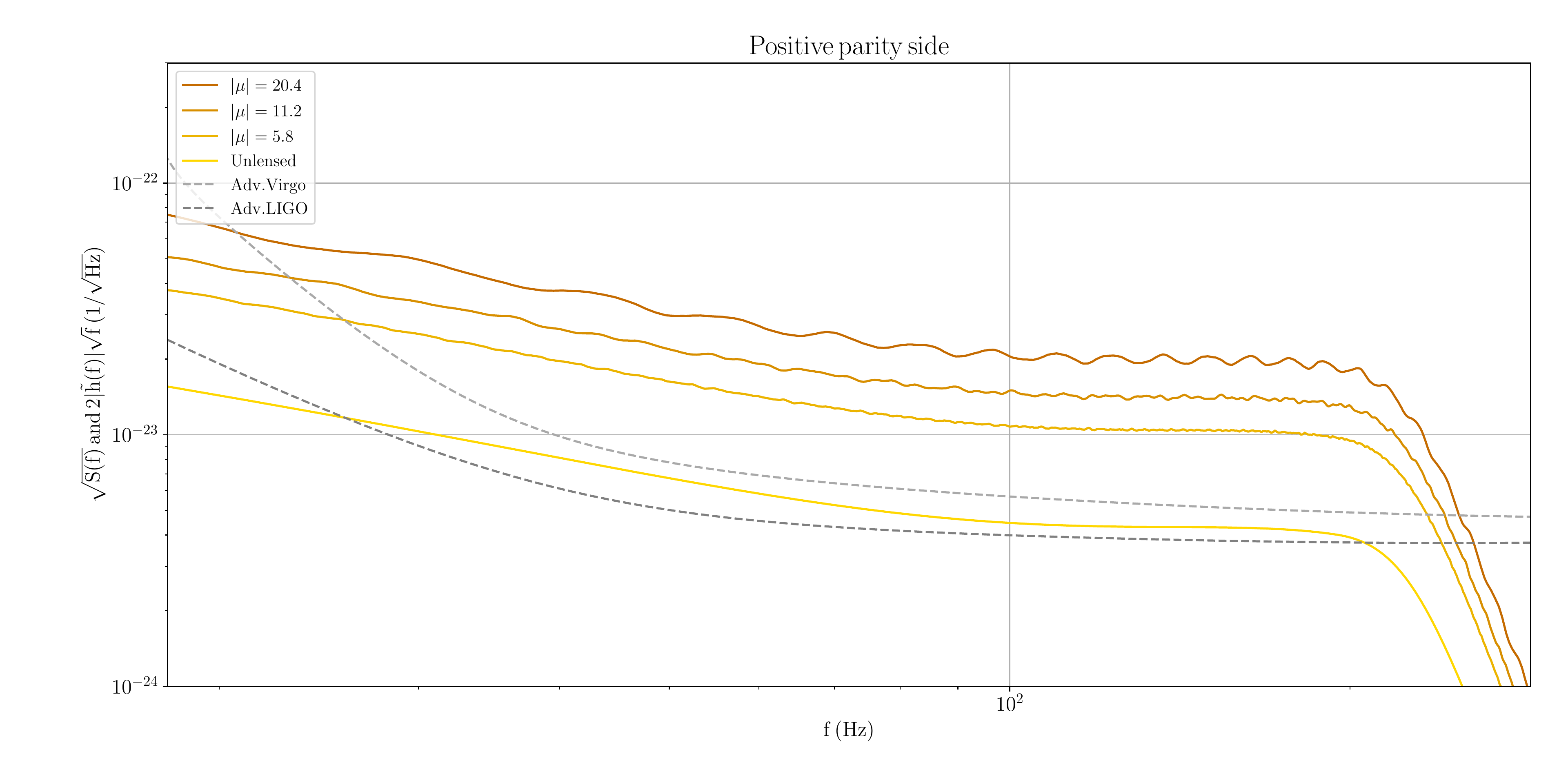}
    \label{fig:elliptical positive parity}
    \end{subfigure}
    
    \begin{subfigure}
    \centering
    \includegraphics[width=1.\textwidth]{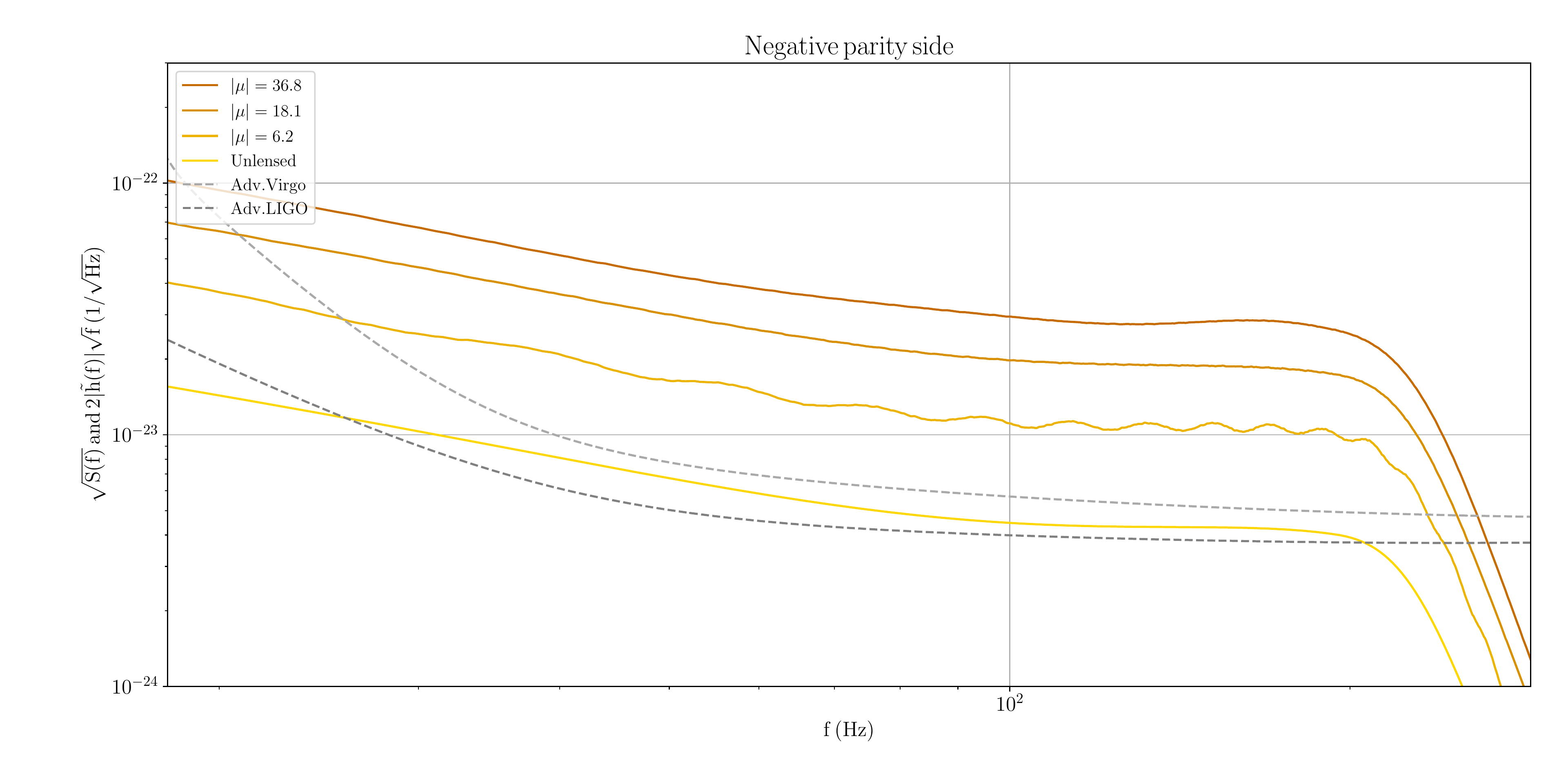}
    \label{fig:elliptical negative parity}
    \end{subfigure}
    
 \caption{\textit{Effects of strongly lensed images' properties on microlensing}. Normalized strains obtained by \textsc{lensingGW} in the Hanford detector for a compact binary lensed by an elliptical galaxy and a microlens background (as documented in Sec.~\ref{subsec:lens equation solver}). Microlensing is enhanced by strong lensing on both the positive parity side (\textit{top panel}) and the negative parity side (\textit{bottom panel}) of the galaxy. As the magnification of the macroimage $|\mu|$ is increased (dark yellow curves, from bottom to top in each panel), the microimages lead to more prominent oscillations in the positive parity side. In contrast, the negative parity side exhibits the opposite behavior. Inspection of the image properties through \textsc{lensingGW} attributes this to the difference in relative magnifications and time delays among the microimages produced by the different scenarios. In both cases, microlensing boosts the SNR of the unlensed signal (light yellow curve) turning it from a sub-threshold event to a detectable one, as demonstrated by the superposed noise curves of LIGO/Virgo (grey lines).}
 \label{fig:macromodel parities}
\end{figure*}

To quantify the effect on gravitational waves, we first solve for the macromodel (the galaxy) only by means of \textsc{lensingGW}'s \texttt{OnlyMacro} option and find the images' magnifications and Morse indices through the dedicated routines. Through those, we identify the macroimage of interest and inject the fixed microlens background around it for each source position. We then solve for the complete model\footnote{ We consider a fixed window of 1.4x1.4\rm{mas} ($\sim 50$ times the Einstein radius of the more massive microlens) for the second step of the iterative procedure.} and compare the lensed and unlensed strains\footnote{The source is a non spinning binary of masses $m_1=45\rm{M_\odot},m_2=36\rm{M_\odot}$ and inclination $\iota =2.6$ at redshift $z_S=2$.} found by \textsc{lensingGW}. We show the strains in the Hanford detector for the positive and negative parity sides in Fig.~\ref{fig:macromodel parities}. 

The unlensed signal is sub-threshold in the Virgo detector and just above threshold in LIGO. However, the amplification of the microimages enhances its amplitude, turning it into a detectable signal on both sides of the galaxy (compare, in particular, the strains to the superposed noise curves of the instruments). The amplitude of the signal increases with the macroimage magnification and the largest time delays between the microimages are of the order of hundreds of milliseconds. This value is larger than the time differences produced by isolated point lenses. The result confirms that increasing the macromodel magnification enhances the effects of microlensing, both in amplitudes and time delays, as found in~\citep{diego2019observational}. 

However, in addition to that, we find that in the positive parity side (\textit{top panel}) the oscillations induced by microlensing become more prominent as the magnification of the strongly lensed image increases. In contrast, the negative parity side (\textit{bottom panel}) exhibits the opposite behavior. This behavior depends on the relative magnifications and time delays of the microimages on the different sides of the galaxy and on the set of microimages which give the most dominant contribution to the gravitational-wave signal. As the effect on the strain is a combination of the various microimage contributions, the lensed waves would be difficult to model analytically. This offers another confirmation of the importance of developing software packages that target arbitrary lensing configurations and related gravitational-wave signals, such as \textsc{lensingGW}.

The execution required $\sim 19 \rm{s}$ on an Intel Core i7-7700HQ CPU @ 2.80GHz for each source position, which makes \textsc{lensingGW} an ideal candidate for systematic studies of such kind.

\section{Conclusions}
\label{sec:conclusions}
It has been proposed that intermediate-mass black holes, dense stellar clusters and primordial black holes could be probed by gravitational-wave lensing in the future~\citep{Jung:2017flg,Lai:2018rto,Christian2018,Diego:2019rzc}. 
Detections of microlensed GWs could also unveil dark matter substructures~\citep{liao2018anomalies}. 
Moreover, if unaccounted for, microlensed signals can introduce biases in the inferred source properties of compact binaries due to the lensed waveform distortions.

We have presented \textsc{lensingGW}, a \textsc{Python} package to predict lensed GWs in ground-based detectors from arbitrary lens models and compact sources.

We have validated \textsc{lensingGW} on two scenarios that produce microlensing: a microlens embedded in a galaxy~\citep{diego2019observational}, where microimages form on top of a strongly lensed image and a binary-point-mass system~\citep{schneiderWeiss}, showing that the software package can recover the correct results.
Moreover, we have demonstrated that the new solving algorithm implemented in \textsc{lensingGW} outperforms standard fixed tile algorithms when applied to hundreds of microlenses embedded in galaxies. The evaluation takes $\sim$ seconds on a standard machine as opposed to several hours required by the fixed tile approach, making \textsc{lensingGW} a valid candidate for investigations of realistic scenarios.

We have shown how to apply \textsc{lensingGW} to investigations of astrophysical interest, such as the impact of the source properties on the detectability of lensed GWs and the effects of strongly lensed images on microlensed strains. Indeed, \textsc{lensingGW} is able to predict lensed GW signals resulting from arbitrary lensing systems such as isolated galaxies or galaxies with hundreds of microlenses thanks to its ability to generate lensed and unlensed gravitational waves.

\section*{Acknowledgements}
We thank Walter Del Pozzo, Gregorio Carullo and Simon Birrer for useful discussions. OAH is supported by the research program of the Netherlands Organization for Scientific Research (NWO). TGFL is partially supported by grants from the Research Grants Council of Hong Kong (Project No. 14306218), Research Committee of the Chinese University of Hong Kong and the Croucher Foundation of Hong Kong. 

\bibliographystyle{aa}

\begin{thebibliography}{75}
	\expandafter\ifx\csname natexlab\endcsname\relax\def\natexlab#1{#1}\fi
	
	\bibitem[{Aasi {et~al.}(2015)}]{TheLIGOScientific:2014jea}
	Aasi, J. {et~al.} 2015, Class. Quantum Grav., 32, 074001
	
	\bibitem[{Abbott {et~al.}(2016)}]{Abbott_2016}
	Abbott, B. {et~al.} 2016, Physical Review Letters, 116
	
	\bibitem[{Abbott {et~al.}(2018{\natexlab{a}})Abbott, Abbott, Abbott, Abernathy,
		Acernese, Ackley, Adams, Adams, Addesso, Adhikari,
		{et~al.}}]{abbott2018:prospects}
	Abbott, B.~P., Abbott, R., Abbott, T., {et~al.} 2018{\natexlab{a}}, Living
	Reviews in Relativity, 21, 3
	
	\bibitem[{{Abbott} {et~al.}(2016){Abbott}, {Abbott}, {Abbott}, {Abernathy},
		{Acernese}, {Ackley}, {Adams}, {Adams}, {Addesso}, {Adhikari},
		{et~al.}}]{2016ApJ...833L...1A}
	{Abbott}, B.~P., {Abbott}, R., {Abbott}, T.~D., {et~al.} 2016, \apjl, 833, L1
	
	\bibitem[{Abbott {et~al.}(2016)}]{TheLIGOScientific:2016agk}
	Abbott, B.~P. {et~al.} 2016, Phys. Rev. Lett., 116, 131103
	
	\bibitem[{Abbott {et~al.}(2017)}]{PhysRevLett.119.161101}
	Abbott, B.~P. {et~al.} 2017, Phys. Rev. Lett., 119, 161101
	
	\bibitem[{Abbott {et~al.}(2018{\natexlab{b}})}]{InstrumentationWhitePaper}
	Abbott, B.~P. {et~al.} 2018{\natexlab{b}},
	\url{https://dcc.ligo.org/LIGO-T1800133}
	
	\bibitem[{Abbott {et~al.}(2019{\natexlab{a}})}]{PhysRevX.9.031040}
	Abbott, B.~P. {et~al.} 2019{\natexlab{a}}, Phys. Rev. X, 9, 031040
	
	\bibitem[{Abbott {et~al.}(2019{\natexlab{b}})}]{PhysRevLett.123.011102}
	Abbott, B.~P. {et~al.} 2019{\natexlab{b}}, Phys. Rev. Lett., 123, 011102
	
	\bibitem[{Abbott {et~al.}(2019{\natexlab{c}})}]{PhysRevD.100.104036}
	Abbott, B.~P. {et~al.} 2019{\natexlab{c}}, Phys. Rev. D, 100, 104036
	
	\bibitem[{{Acernese, F. and others}(2015)}]{TheVirgo:2014hva}
	{Acernese, F. and others}. 2015, Class. Quantum Grav., 32, 024001
	
	\bibitem[{Akutsu {et~al.}(2018)Akutsu, Ando, Araki, Araya, Arima, Aritomi,
		Asada, Aso, Atsuta, Awai, {et~al.}}]{akutsu2018construction}
	Akutsu, T., Ando, M., Araki, S., {et~al.} 2018, Progress of Theoretical and
	Experimental Physics, 2018, 013F01
	
	\bibitem[{Allen {et~al.}(2012)Allen, Anderson, Brady, Brown, \&
		Creighton}]{Allen_2012}
	Allen, B., Anderson, W.~G., Brady, P.~R., Brown, D.~A., \& Creighton, J. D.~E.
	2012, Physical Review D, 85
	
	\bibitem[{Aso {et~al.}(2013)Aso, Michimura, Somiya, Ando, Miyakawa, Sekiguchi,
		Tatsumi, Yamamoto, Collaboration, {et~al.}}]{aso2013interferometer}
	Aso, Y., Michimura, Y., Somiya, K., {et~al.} 2013, Physical Review D, 88,
	043007
	
	\bibitem[{Baker \& Trodden(2017)}]{PhysRevD.95.063512}
	Baker, T. \& Trodden, M. 2017, Phys. Rev. D, 95, 063512
	
	\bibitem[{{Birrer} \& {Amara}(2018)}]{lenstronomy}
	{Birrer}, S. \& {Amara}, A. 2018, {Lenstronomy: Multi-purpose gravitational
		lens modeling software package}, Astrophysics Source Code Library
	
	\bibitem[{Bliokh \& Minakov(1975)}]{bliokh1975diffraction}
	Bliokh, P. \& Minakov, A. 1975, Astrophysics and Space Science, 34, L7
	
	\bibitem[{Bontz \& Haugan(1981)}]{bontz1981diffraction}
	Bontz, R.~J. \& Haugan, M.~P. 1981, Astrophysics and Space Science, 78, 199
	
	\bibitem[{Broadhurst {et~al.}(2019)Broadhurst, Diego, \&
		Smoot}]{Broadhurst:2019ijv}
	Broadhurst, T., Diego, J.~M., \& Smoot, G.~F. 2019 [\eprint[arXiv]{1901.03190}]
	
	\bibitem[{Broadhurst {et~al.}(2018)Broadhurst, Diego, \&
		Smoot~III}]{broadhurst2018reinterpreting}
	Broadhurst, T., Diego, J.~M., \& Smoot~III, G. 2018, arXiv preprint
	arXiv:1802.05273
	
	\bibitem[{Cao {et~al.}(2019)Cao, Qi, Cao, Biesiada, Li, Pan, \&
		Zhu}]{Cao:2019kgn}
	Cao, S., Qi, J., Cao, Z., {et~al.} 2019, Sci. Rep., 9, 11608
	
	\bibitem[{Cao {et~al.}(2014)Cao, Li, \& Wang}]{cao2014gravitational}
	Cao, Z., Li, L.-F., \& Wang, Y. 2014, Physical Review D, 90, 062003
	
	\bibitem[{{Christian} {et~al.}(2018){Christian}, {Vitale}, \&
		{Loeb}}]{Christian2018}
	{Christian}, P., {Vitale}, S., \& {Loeb}, A. 2018, \prd, 98, 103022
	
	\bibitem[{{Collett}(2015)}]{collett2015}
	{Collett}, T.~E. 2015, \apj, 811, 20
	
	\bibitem[{Collett \& Bacon(2017)}]{PhysRevLett.118.091101}
	Collett, T.~E. \& Bacon, D. 2017, Phys. Rev. Lett., 118, 091101
	
	\bibitem[{Contigiani(2020)}]{Contigiani:2020yyc}
	Contigiani, O. 2020, Monthly Notices of the Royal Astronomical Society, 492,
	3359
	
	\bibitem[{Cutler \& Flanagan(1994)}]{cutlerFlanagan}
	Cutler, C. \& Flanagan, E.~E. 1994, Phys. Rev. D, 49, 2658
	
	\bibitem[{Dai {et~al.}(2018)Dai, Li, Zackay, Mao, \& Lu}]{dai2018detecting}
	Dai, L., Li, S.-S., Zackay, B., Mao, S., \& Lu, Y. 2018, Physical Review D, 98,
	104029
	
	\bibitem[{Dai \& Venumadhav(2017)}]{Dai:2017huk}
	Dai, L. \& Venumadhav, T. 2017 [\eprint[arXiv]{1702.04724}]
	
	\bibitem[{Dai {et~al.}(2017)Dai, Venumadhav, \& Sigurdson}]{dai2017effect}
	Dai, L., Venumadhav, T., \& Sigurdson, K. 2017, Physical Review D, 95, 044011
	
	\bibitem[{Deguchi \& Watson(1986)}]{deguchi1986diffraction}
	Deguchi, S. \& Watson, W. 1986, The Astrophysical Journal, 307, 30
	
	\bibitem[{Diego(2019{\natexlab{a}})}]{Diego:2019rzc}
	Diego, J.~M. 2019{\natexlab{a}} [\eprint[arXiv]{1911.05736}]
	
	\bibitem[{Diego(2019{\natexlab{b}})}]{Diego:2018fzr}
	Diego, J.~M. 2019{\natexlab{b}}, Astron. Astrophys., 625, A84
	
	\bibitem[{Diego {et~al.}(2017)Diego, Kaiser, Broadhurst, Kelly, Rodney,
		Morishita, Oguri, Ross, Zitrin, Jauzac, {et~al.}}]{diego2017dark}
	Diego, J.~M., Kaiser, N., Broadhurst, T., {et~al.} 2017, arXiv preprint
	arXiv:1706.10281
	
	\bibitem[{Diego {et~al.}(2018)Diego, Kaiser, Broadhurst, Kelly, Rodney,
		Morishita, Oguri, Ross, Zitrin, Jauzac, {et~al.}}]{diego2018dark}
	Diego, J.~M., Kaiser, N., Broadhurst, T., {et~al.} 2018, The Astrophysical
	Journal, 857, 25
	
	\bibitem[{{Diego, J. M.} {et~al.}(2019){Diego, J. M.}, {Hannuksela, O. A.},
		{Kelly, P. L.}, {Pagano, G.}, {Broadhurst, T.}, {Kim, K.}, {Li, T. G. F.}, \&
		{Smoot, G. F.}}]{diego2019observational}
	{Diego, J. M.}, {Hannuksela, O. A.}, {Kelly, P. L.}, {et~al.} 2019, A\&A, 627,
	A130
	
	\bibitem[{Fan {et~al.}(2017)Fan, Liao, Biesiada, Pi\'orkowska-Kurpas, \&
		Zhu}]{PhysRevLett.118.091102}
	Fan, X.-L., Liao, K., Biesiada, M., Pi\'orkowska-Kurpas, A., \& Zhu, Z.-H.
	2017, Phys. Rev. Lett., 118, 091102
	
	\bibitem[{Hannuksela {et~al.}(2019)Hannuksela, Haris, Ng, Kumar, Mehta, Keitel,
		Li, \& Ajith}]{hannuksela2019search}
	Hannuksela, O., Haris, K., Ng, K., {et~al.} 2019, The Astrophysical Journal
	Letters, 874, L2
	
	\bibitem[{Hannuksela {et~al.}(2020)Hannuksela, Collett, Çal\i~\c skan, \&
		Li}]{Hannuksela:2020xor}
	Hannuksela, O.~A., Collett, T.~E., Çal\i~\c skan, M., \& Li, T.~G. 2020
	[\eprint[arXiv]{2004.13811}]
	
	\bibitem[{Haris {et~al.}(2018)Haris, Mehta, Kumar, Venumadhav, \&
		Ajith}]{Haris:2018vmn}
	Haris, K., Mehta, A.~K., Kumar, S., Venumadhav, T., \& Ajith, P. 2018
	[\eprint[arXiv]{1807.07062}]
	
	\bibitem[{Hou {et~al.}(2020)Hou, Fan, Liao, \& Zhu}]{Hou:2019dcm}
	Hou, S., Fan, X.-L., Liao, K., \& Zhu, Z.-H. 2020, Phys. Rev. D, 101, 064011
	
	\bibitem[{{Iyer} {et~al.}(2011)}]{M1100296}
	{Iyer}, B. {et~al.} 2011, {LIGO India}, Tech. Rep. LIGO-M1100296,
	https://dcc.ligo.org/LIGO-M1100296/public
	
	\bibitem[{Jung \& Shin(2019)}]{Jung:2017flg}
	Jung, S. \& Shin, C.~S. 2019, Phys. Rev. Lett., 122, 041103
	
	\bibitem[{{Keeton}(2011)}]{gravlens}
	{Keeton}, C.~R. 2011, {GRAVLENS: Computational Methods for Gravitational
		Lensing}
	
	\bibitem[{Lai {et~al.}(2018)Lai, Hannuksela, Herrera-Mart{\'\i}n, Diego,
		Broadhurst, \& Li}]{Lai:2018rto}
	Lai, K.-H., Hannuksela, O.~A., Herrera-Mart{\'\i}n, A., {et~al.} 2018, Phys.
	Rev., D98, 083005
	
	\bibitem[{{Li} {et~al.}(2019){Li}, {Lo}, {Sachdev}, {Chan}, {Lin}, {Li}, \&
		{Weinstein}}]{2019arXiv190406020L}
	{Li}, A. K.~Y., {Lo}, R. K.~L., {Sachdev}, S., {et~al.} 2019, arXiv e-prints,
	arXiv:1904.06020
	
	\bibitem[{Li {et~al.}(2018)Li, Mao, Zhao, \& Lu}]{li2018gravitational}
	Li, S.-S., Mao, S., Zhao, Y., \& Lu, Y. 2018, Monthly Notices of the Royal
	Astronomical Society, 476, 2220
	
	\bibitem[{Liao {et~al.}(2018)Liao, Ding, Biesiada, Fan, \&
		Zhu}]{liao2018anomalies}
	Liao, K., Ding, X., Biesiada, M., Fan, X.-L., \& Zhu, Z.-H. 2018, The
	Astrophysical Journal, 867, 69
	
	\bibitem[{Liao {et~al.}(2017)Liao, Fan, Ding, Biesiada, \& Zhu}]{Liao:2017ioi}
	Liao, K., Fan, X.-L., Ding, X.-H., Biesiada, M., \& Zhu, Z.-H. 2017, Nature
	Commun., 8, 1148, [Erratum: Nature Commun. 8, 2136 (2017)]
	
	\bibitem[{{LIGO Scientific Collaboration}(2018)}]{lalsuite}
	{LIGO Scientific Collaboration}. 2018, {LIGO} {A}lgorithm {L}ibrary -
	{LALS}uite, free software (GPL)
	
	\bibitem[{{LIGO Scientific Collaboration} \& {Virgo
			Collaboration}(2019)}]{GCNO3}
	{LIGO Scientific Collaboration} \& {Virgo Collaboration}. 2019,
	\url{https://gcn.gsfc.nasa.gov/gcn3/24045.gcn3}
	
	\bibitem[{{McIsaac} {et~al.}(2019){McIsaac}, {Keitel}, {Collett}, {Harry},
		{Mozzon}, {Edy}, \& {Bacon}}]{2019arXiv191205389M}
	{McIsaac}, C., {Keitel}, D., {Collett}, T., {et~al.} 2019, arXiv e-prints,
	arXiv:1912.05389
	
	\bibitem[{Mehta {et~al.}(2019)Mehta, Haris, Venumadhav, \&
		Ajith}]{Mehta:2019aa}
	Mehta, A.~K., Haris, K., Venumadhav, T., \& Ajith, P. 2019, in preparation
	
	\bibitem[{{Mukherjee} {et~al.}(2019){Mukherjee}, {Wandelt}, \&
		{Silk}}]{2019arXiv190808950M}
	{Mukherjee}, S., {Wandelt}, B.~D., \& {Silk}, J. 2019, arXiv e-prints,
	arXiv:1908.08950
	
	\bibitem[{Nakamura(1998)}]{nakamura1998gravitational}
	Nakamura, T.~T. 1998, Physical review letters, 80, 1138
	
	\bibitem[{Ng {et~al.}(2018)Ng, Wong, Broadhurst, \& Li}]{ng2017precise}
	Ng, K.~K., Wong, K.~W., Broadhurst, T., \& Li, T.~G. 2018, Physical Review D,
	97, 023012
	
	\bibitem[{{Oguri}(2018)}]{2018MNRAS.480.3842O}
	{Oguri}, M. 2018, \mnras, 480, 3842
	
	\bibitem[{{Oguri}(2019)}]{2019RPPh...82l6901O}
	{Oguri}, M. 2019, Reports on Progress in Physics, 82, 126901
	
	\bibitem[{Ohanian(1974)}]{ohanian1974focusing}
	Ohanian, H.~C. 1974, International Journal of Theoretical Physics, 9, 425
	
	\bibitem[{{Pagano} {et~al.}(in prep.){Pagano}, {Hannuksela}, \&
		{Li}}]{giuliaforwardcitation}
	{Pagano}, G., {Hannuksela}, O., \& {Li}, T. G.~F. in prep.
	
	\bibitem[{Pang {et~al.}(2020)Pang, Hannuksela, Dietrich, Pagano, \&
		Harry}]{10.1093/mnras/staa1430}
	Pang, P.~T.~H., Hannuksela, O.~A., Dietrich, T., Pagano, G., \& Harry, I.~W.
	2020, Monthly Notices of the Royal Astronomical Society, 495, 3740
	
	\bibitem[{Press {et~al.}(2007)Press, Teukolsky, Vetterling, \&
		Flannery~Frontmatt}]{press1992}
	Press, W.~H., Teukolsky, S.~A., Vetterling, W.~T., \& Flannery~Frontmatt, B.~P.
	2007, {NUMERICAL RECIPES}, 3rd edn. (Cambridge: {Cambridge University Press})
	
	\bibitem[{{Robertson} {et~al.}(2020){Robertson}, {Smith}, {Massey}, {Eke},
		{Jauzac}, {Bianconi}, \& {Ryczanowski}}]{2020arXiv200201479R}
	{Robertson}, A., {Smith}, G.~P., {Massey}, R., {et~al.} 2020, arXiv e-prints,
	arXiv:2002.01479
	
	\bibitem[{Schneider {et~al.}(2006)Schneider, Kochanek, \&
		Wambsganss}]{schneiderBook}
	Schneider, P., Kochanek, C., \& Wambsganss, J. 2006, {G}ravitational {L}ensing:
	{S}trong, {W}eak and {M}icro, ed. G.~Meylan, P.~Jetzer, \& P.~North (Berlin:
	Springer-Verlag Berlin Heidelberg)
	
	\bibitem[{Schneider \& Weiss(1986)}]{schneiderWeiss}
	Schneider, P. \& Weiss, A. 1986, Astron. Astrophys., 164, 237–259
	
	\bibitem[{{Sereno} {et~al.}(2011){Sereno}, {Jetzer}, {Sesana}, \&
		{Volonteri}}]{2011MNRAS.415.2773S}
	{Sereno}, M., {Jetzer}, P., {Sesana}, A., \& {Volonteri}, M. 2011, \mnras, 415,
	2773
	
	\bibitem[{Singer {et~al.}(2019)Singer, Goldstein, \& Bloom}]{Singer:2019vjs}
	Singer, L.~P., Goldstein, D.~A., \& Bloom, J.~S. 2019
	[\eprint[arXiv]{1910.03601}]
	
	\bibitem[{{Smith} {et~al.}(2018){Smith}, {Berry}, {Bianconi}, {Farr}, {Jauzac},
		{Massey}, {Richard}, {Robertson}, {Sharon}, {Vecchio}, \&
		{Veitch}}]{2018IAUS..338...98S}
	{Smith}, G.~P., {Berry}, C., {Bianconi}, M., {et~al.} 2018, in IAU Symposium,
	Vol. 338, IAU Symposium, ed. G.~{Gonz{\'a}lez} \& R.~{Hynes}, 98--102
	
	\bibitem[{{Smith} {et~al.}(2019){Smith}, {Bianconi}, {Jauzac}, {Richard},
		{Robertson}, {Berry}, {Massey}, {Sharon}, {Farr}, \&
		{Veitch}}]{2019MNRAS.485.5180S}
	{Smith}, G.~P., {Bianconi}, M., {Jauzac}, M., {et~al.} 2019, \mnras, 485, 5180
	
	\bibitem[{Somiya(2012)}]{somiya:2012detector}
	Somiya, K. 2012, Classical and Quantum Gravity, 29, 124007
	
	\bibitem[{Sun \& Fan(2019)}]{Sun:2019ztn}
	Sun, D. \& Fan, X. 2019 [\eprint[arXiv]{1911.08268}]
	
	\bibitem[{Takahashi \&
		Nakamura(2003{\natexlab{a}})}]{takahashi2003gravitational}
	Takahashi, R. \& Nakamura, T. 2003{\natexlab{a}}, The Astrophysical Journal,
	595, 1039
	
	\bibitem[{Takahashi \& Nakamura(2003{\natexlab{b}})}]{takahashi2003wave}
	Takahashi, R. \& Nakamura, T. 2003{\natexlab{b}}, The Astrophysical Journal,
	595, 1039
	
	\bibitem[{Takahashi {et~al.}(2005)Takahashi, Suyama, \&
		Michikoshi}]{takahashi2005scattering}
	Takahashi, R., Suyama, T., \& Michikoshi, S. 2005, Astronomy \& Astrophysics,
	438, L5
	
	\bibitem[{Thorne(1983)}]{thorne1983theory}
	Thorne, K.~S. 1983, in Gravitational radiation, 1--57
	
\end{thebibliography}
 \newcommand{\noop}[1]{}

\end{document}